\documentclass[fleqn,usenatbib,useAMS,onecolumn]{mnras}

\usepackage[T1]{fontenc}
\usepackage{ae,aecompl}

\usepackage{newtxtext,newtxmath}

\usepackage[british]{babel}
\usepackage{graphicx}
\usepackage{subfig}
\usepackage{amsmath}

\setcitestyle{notesep={}}

\title[The \textit{r}-modes of stratified neutron stars]{The \textit{r}-modes of slowly rotating, stratified neutron stars}

\author[F. Gittins and N. Andersson]{Fabian Gittins\thanks{E-mail: f.w.r.gittins@soton.ac.uk (FG); n.a.andersson@soton.ac.uk (NA)} and Nils Andersson\footnotemark[1] \\
Mathematical Sciences and STAG Research Centre, University of Southampton, Southampton SO17 1BJ, United Kingdom}

\date{}

\pubyear{2023}

\begin{document}
\label{firstpage}
\pagerange{\pageref{firstpage}--\pageref{lastpage}}
\maketitle

\begin{abstract}
The only \textit{r}-modes that exist in a globally barotropic, rotating, Newtonian star are the fundamental $l = |m|$ solutions, where $l$ and $m$ are the indices of the spherical harmonic $Y_l^m$ that describe the mode's angular dependence. This is in stark contrast to a stellar model that is non-barotropic throughout its interior, which hosts all the $l \geq |m|$ perturbations including radial overtones. In reality, neutron stars are stratified with locally barotropic regions. Therefore, we explore how stratification alters a star's ability to support \textit{r}-modes. We consider the globally stratified case and examine the behaviour of the modes as the star gets close to barotropicity. In this limit, we find that all but the fundamental $l = |m|$ perturbations change character and become generic inertial modes. Restricting the analysis to $l = |m|$ perturbations, we develop the \textit{r}-mode equations in order to consider stellar models that exhibit local barotropicity. Our results for such models show that the \textit{r}-mode overtones diverge and join the inertial modes. In order to see which \textit{r}-modes persist and retain their character in realistic neutron stars, these calculations will need to be brought into full general relativity.
\end{abstract}



\section{Introduction}

Stellar interiors host a rich spectrum of oscillation modes \citep{1980tsp..book.....C,1989nos..book.....U}. These perturbations are sensitive to the characteristics of the star. Indeed, each ingredient of physics -- density, stratification, rotation, magnetic field etc. -- corresponds (more or less) directly to a unique set of oscillation modes. It would seem, however, that not all oscillation modes are created equal. An interesting family of modes that arise due to stellar rotation are the \textit{r}-modes. These Coriolis-dominated perturbations possess the remarkable property of being generically unstable in a perfect-fluid star due to the emission of gravitational radiation \citep{1998ApJ...502..708A,1998ApJ...502..714F}. This result inspired a wide body of literature \citep[see reviews][]{2001IJMPD..10..381A,2003CQGra..20R.105A}, including suggestions that the \textit{r}-mode instability may limit the rotation rates of newly born pulsars \citep{1998PhRvL..80.4843L,1999ApJ...510..846A} and more mature accreting systems \citep{1998ApJ...501L..89B,1999ApJ...516..307A}. The \textit{r}-modes are also candidates for gravitational-wave observations, with a recent search focused on the glitching neutron star PSR J0537-6910 \citep{2021ApJ...922...71A}. However, so far no gravitational-wave signatures consistent with an \textit{r}-mode instability have been seen.

It was \citet{1941MNRAS.101..367C}, in his work on non-rotating polytropes, who provided the first classification of modes according to the physics dominating their behaviour. The simplest stellar model is that of a spherical, incompressible fluid. Such a star will only have one family of oscillation modes, the fundamental \textit{f}-modes. The \textit{f}-modes are distinguished by having no nodes in their radial eigenfunctions and inducing large density perturbations in the star. Should one allow for compressibility of the stellar fluid, by including an equation of state, then the \textit{p}-modes arise. These are high-frequency acoustic waves restored by the pressure of the fluid, also associated with large perturbations in the density. Suppose the star becomes stratified such that the matter is no longer barotropic.%
\footnote{A star is said to be barotropic if the equilibrium and perturbed configurations satisfy the same one-parameter equation of state. Earlier work instead referred to such a star as ``isentropic'', because an isentropic star with no composition gradients has this property. However, for neutron stars, the departure from a one-parameter equation of state is dominated by composition gradients. Thus, an isentropic neutron star may not, in general, have the same one-parameter equation of state for the background and the perturbation.}
Then \textit{g}-modes will appear, restored by gravity gradients. The \textit{g}-modes have low frequencies and small density perturbations.

These three families of fluid oscillations, the \textit{f}-, \textit{p}- and \textit{g}-modes, all belong to the class of polar modes. A perturbation is polar (spheroidal or even parity) if it varies like a spherical harmonic $Y_l^m$ under a parity transformation. In a spherically symmetric, fluid star, this is the only class of modes that exists. Furthermore, the perturbations of a spherical star are especially simple as each mode is associated with a single $Y_l^m$. The class of axial (toroidal or odd parity) modes, which transform opposite under parity to polar modes, require some level of anisotropy in order to exist.

When the star begins to rotate, a new family of modes appears. These are known as the inertial modes and they are restored by the Coriolis force \citep{1889RSPTA.180..187B,greenspan,1999PhRvD..59d4009L,2001ApJ...550..443R}. Even in the non-rotating limit, the inertial modes generically involve a mix of polar and axial perturbations with definite parity and couple multiple spherical harmonics \citep{1992ApJ...397..674L,1999ApJ...521..764L,2000ApJ...529..997Y}. Formally, these modes inhabit a zero-frequency subspace on the spherical star, being stationary convective fluid currents, and attain non-zero oscillation frequencies at first order in the star's angular frequency. Among these modes exists a special subclass, the purely axial \textit{r}-modes.

The \textit{r}-modes were first studied in the context of astrophysics by \citet{1978MNRAS.182..423P}, who named them for their similarity to the Rossby waves of terrestrial meteorology \citep[see][ for a recent review]{2021SSRv..217...15Z}. Contrary to a typical inertial mode, an \textit{r}-mode is an axial perturbation associated with a single $Y_l^m$ on the spherical star. Moreover in Newtonian gravity, their leading-order frequency in a slow-rotation expansion can be determined analytically to be 
\begin{equation}
    \omega_0 = \frac{2 m}{l (l + 1)} \Omega
    \label{eq:omega_0}
\end{equation}
in the rotating frame of the star with angular frequency $\Omega$. Since the frequency of a mode as measured by an inertial observer outside the star is simply $\omega_0 - m \Omega$, the \textit{r}-modes are retrograde in the frame of the star, but prograde in the inertial frame at all rates of rotation. Therefore, they satisfy the well-known Chandrasekhar-Friedman-Schutz instability criterion and are generically unstable to perturbations driven by gravitational radiation \citep{1970PhRvL..24..611C,1978ApJ...222..281F,1998ApJ...502..708A,1998ApJ...502..714F}.

To date, most of the \textit{r}-mode calculations have involved Newtonian stellar models that are globally barotropic \citep{1998PhRvL..80.4843L,1999ApJ...521..764L,2000ApJ...529..997Y} or globally non-barotropic \citep{1981A&A....94..126P,1981Ap&SS..78..483S,1982ApJ...256..717S,1999ApJ...510..846A,2000ApJS..129..353Y}. The barotropic star is especially simple: there exist only the fundamental $l = |m|$ \textit{r}-modes in the star and the leading-order axial eigenfunctions can be obtained analytically, independent of the equation of state. However, the non-barotropic case is more complicated. One must work to beyond leading order in rotation, where the star departs from spherical symmetry \citep{1978MNRAS.182..423P,1981A&A....94..126P,1982ApJ...256..717S}. The equations then admit a standard eigenvalue problem for the $l \geq |m|$ \textit{r}-mode solutions. The reason for the concentration on Newtonian stars has been in part due to the challenges in calculating the \textit{r}-modes in general relativity.

Our particular focus is on neutron stars, which are highly relativistic bodies. In order to involve a realistic description for the nuclear matter, we need to formulate the \textit{r}-mode problem beyond Newtonian gravity. In relativity, there are no longer any purely axial inertial modes in barotropes \citep[except for stationary dipole perturbations;][]{1999PhDT........10L}. In this direction, there have been a number of calculations of the relativistic inertial modes \citep{2000PhRvD..63b4019L,2003PhRvD..68l4010L,2003MNRAS.339.1170R}, including physically motivated equations of state \citep{2015PhRvD..91b4001I}. The relativistic inertial modes may or may not be a reasonable approximation of the problem. In reality, neutron stars are stratified due to varying chemical composition \citep{1992ApJ...395..240R,2019MNRAS.489.4043A}. The \textit{r}-modes exist in non-barotropic, relativistic stars. However, the relativistic perturbation equations imply a continuous spectrum \citep{1998MNRAS.293...49K,1999MNRAS.308..745B}. This is surprising, since the \textit{r}-modes have well-defined frequencies in Newtonian gravity. It is unclear whether the continuous spectrum is physical or an artefact of some simplifying assumptions. Adopting the latter view, there have been efforts to regularise the problem \citep{2004CQGra..21.4661L,2005MNRAS.363..121P}, as well as studies using the Cowling approximation \citep{1999ApJ...520..788K,2022Univ....8..542K,2022PhRvD.106j3009K}.

In this paper, our goal is to study the role of stratification for the \textit{r}-modes of Newtonian stars. In particular, we want to move beyond global assumptions about the matter and consider the more realistic case where the fluid may be locally barotropic. This is expected to be the case for neutron stars; their high-density cores will likely have composition gradients, whereas the outer layers will be barotropic (since matter at low densities is composed of single nuclei). Our hope is that, in understanding the Newtonian problem, we may make progress towards calculating the relativistic \textit{r}-modes.

This paper is organised as follows. We begin in Section~\ref{sec:background} with a brief discussion on constructing slowly rotating stellar models, which will form the background for the \textit{r}-mode oscillations. We move on to Section~\ref{sec:perturbations} to describe the perturbation formalism for rotating stars and present the \textit{r}-mode equations. We develop these equations into an eigenvalue problem in Section~\ref{sec:non-barotropic}, assuming that the star is globally non-barotropic. These expressions are used to calculate the \textit{r}-modes as the star gets close to the barotropic limit. In Section~\ref{sec:locally-barotropic}, seeking to consider more realistic stellar models, we derive the $l = |m|$ system of equations, where the matter may be locally barotropic. We implement some neutron-star equations of state for the perturbations with a polytropic background and compute the oscillations. Finally, we summarise and suggest future directions in Section~\ref{sec:conclusions}.

\section{Slowly rotating background}
\label{sec:background}

We will examine the \textit{r}-modes of a slowly rotating, Newtonian star. In principle, this limits the rates of rotation our analysis is accurate to. However, this will enable us to explore the character of the modes and, in practice, many stars fall comfortably within the slow-rotation regime.

The structure of a uniformly rotating star is a solution to the following system of equations:
\begin{subequations}\label{eqs:rotating}
\begin{gather}
    \frac{1}{\rho} \nabla_j p = - \nabla_j \Psi, \label{eq:rotatingEuler}\\
    \nabla_j \nabla^j \Phi = 4 \uppi G \rho,
\end{gather}
with an equation of state 
\begin{equation}
    p = p(\rho),
\end{equation}
\end{subequations}
where 
\begin{equation}
    \Psi = \Phi - \frac{1}{3} \Omega^2 r^2 [1 - P_2(\cos \theta)]
    \label{eq:effectivePotential}
\end{equation}
is the effective potential -- the sum of the gravitational $\Phi$ and centrifugal potentials -- $\rho$ is the mass density and $p$ is the pressure. Here, we use spherical polar coordinates $(r, \theta, \phi)$, where the star rotates with angular velocity $\Omega^j$ about the $\theta = 0$ axis, $P_l$ is a Legendre polynomial of degree $l$ and $\nabla_j$ is the covariant derivative. We assume the star to be rotating slowly such that 
\begin{equation}
    1 \gg \epsilon^2 \equiv \frac{\Omega^2}{G M / R^3},
    \label{eq:slowRotation}
\end{equation}
where $M$ is the total mass of the star and $R$ is the radius of its corresponding spherical configuration.%
\footnote{The corresponding spherical star must be the same star when $\Omega = 0$. In order for it to be the same star in any meaningful sense, the two configurations must have identical masses and be described by the same equation of state. Consequently, this means that the rotating configuration will have a smaller central mass density, as fluid elements will move away from the centre due to the centrifugal force.}
In this context, a perturbative approach to solving equations~\eqref{eqs:rotating} is permitted, where the departure from sphericity is small.%
\footnote{We note that the fastest observed pulsar to date has a recorded spin frequency of $716 \ \text{Hz}$ \citep{2006Sci...311.1901H}. Assuming a canonical $M = 1.4 \ \text{M}_\odot$, $R = 10 \ \text{km}$ neutron star, this corresponds to $\epsilon^2 \approx 0.1$, which can be reasonably argued to fall under slow rotation.}

In a spherical star, the surfaces of constant density, pressure and gravitational potential coincide and depend solely on $r$. This is not so when the star begins to rotate, as the centrifugal force spoils the symmetry with respect to the polar angle $\theta$. Therefore, in calculating rotating configurations, it is convenient to adjust the coordinate system in the following way. We use the Clairaut-Legendre expansion \citep[see, e.g., Section~5.2 of][]{1978trs..book.....T} and introduce a new coordinate $a$ that corresponds to the isobaric surfaces such that $\rho(r, \theta)= \rho(a)$ (and thus also corresponds to the isopycnic surfaces). By the Euler equation~\eqref{eq:rotatingEuler}, these surfaces will also coincide with the surfaces of constant $\Psi$. Since the star rotates slowly, we may define 
\begin{equation}
    r(a, \theta) = a [1 + \varepsilon(a, \theta)] + O(\epsilon^4),
    \label{eq:coordinate}
\end{equation}
where $\varepsilon = O(\epsilon^2)$ characterises the deviation from spherical symmetry. (That the expansion only involves even powers of $\epsilon$ is a consequence of the rotational symmetry.) This coordinate change is associated with the metric tensor $g_{j k}$ defined by the line element 
\begin{equation}
    ds^2 = g_{j k} dx^j dx^k = (1 + 2 \varepsilon) [da^2 + a^2 (d\theta^2 + \sin^2 \theta \, d\phi^2)] + 2 a (\partial_a \varepsilon \, da + \partial_\theta \varepsilon \, d\theta) da + O(\epsilon^4).
\end{equation}
We note that the metric differs from the usual spherical polar coordinates at $O(\epsilon^2)$ and the coordinate basis is no longer orthogonal. We can see from \eqref{eq:effectivePotential} that the centrifugal potential only has contributions from the $l = 0, 2$ Legendre polynomials. Hence, the problem decouples into these two sectors and we must have 
\begin{equation}
    \varepsilon(a, \theta) = \varepsilon_0(a) + \varepsilon_2(a) P_2(\cos \theta).
\end{equation}

The $l = 2$ sector characterises the shape of the star and reduces to Clairaut's equation 
\begin{equation}
    a^2 \frac{d^2 \varepsilon_2}{da^2} + 2 \frac{d \ln m_0}{d \ln a} \frac{d (a \varepsilon_2)}{da} = 6 \varepsilon_2,
    \label{eq:shape}
\end{equation}
where $m_0 = O(1)$ denotes the mass distribution of the non-rotating configuration. We also note $\rho_0 = O(1)$, $p_0 = O(1)$ and $\Phi_0 = O(1)$ as the mass density, pressure and gravitational potential of the spherical star, respectively. An examination of the behaviour at the centre shows that $\varepsilon_2$ approaches a constant. The second boundary condition comes from matching the interior solution for $\Phi$ to the exterior solution that decays as $1 / r^3$. Therefore, we find 
\begin{equation}
    2 \varepsilon_2(R) + R \left. \frac{d \varepsilon_2}{da} \right|_{a = R} = - \frac{5}{3} \epsilon^2.
\end{equation}
Equipped with the non-spherical shape of the slowly rotating background to second order in rotation, we go on to formulate the perturbation problem.

\section{The perturbations}
\label{sec:perturbations}

Oscillation modes are harmonic solutions to the following linearised equations of motion (expressed in the rotating frame):
\begin{subequations}\label{eqs:perturbed}
\begin{gather}
    \delta \rho = - \nabla_j (\rho \xi^j), \\
    \partial_t^2 \xi_j + 2 \epsilon_{j k n} \Omega^k \partial_t \xi^n = - \frac{1}{\rho} \nabla_j \delta p + \frac{\delta \rho}{\rho^2} \nabla_j p - \nabla_j \delta \Phi, \label{eq:perturbedEuler}\\
    \nabla_j \nabla^j \delta \Phi = 4 \uppi G \delta \rho,
\end{gather}
with an equation of state for the perturbations 
\begin{equation}
    \frac{\Delta p}{p} = \Gamma_1 \frac{\Delta \rho}{\rho},
    \label{eq:perturbedEOS}
\end{equation}
\end{subequations}
where $\delta$ and $\Delta$ denote the Eulerian and Lagrangian variations of a quantity, respectively, $\xi^j$ is the Lagrangian displacement vector of the fluid elements and $\Gamma_1 = \Gamma_1(a)$ is the adiabatic index of the perturbations. The stratification of the fluid enters the description through the linearised equation of state~\eqref{eq:perturbedEOS}. As a starting assumption, it is common to choose $\Gamma_1 = \text{const}$. This leads to qualitative insight, but one has to be careful in drawing quantitative conclusions. In general, when $\Gamma_1$ differs from the adiabatic index of the background
\begin{equation}
    \Gamma = \frac{d \ln p}{d \ln \rho},
\end{equation}
the perturbations obey a different equation of state and thus the star is non-barotropic.

Since the equilibrium configuration is axisymmetric with respect to the azimuthal coordinate $\phi$, each mode will have a definite order $m$ and we may assume the following form for the Lagrangian displacement in our $(a, \theta, \phi)$ coordinate basis:
\begin{subequations}\label{eqs:displacement}
\begin{align}
    \xi^a &= \frac{1}{a} \sum_{l = |m|}^\infty W_l(a) Y_l^m e^{i \omega t}, \\
    \xi^\theta &= \frac{1}{a^2} \sum_{l = |m|}^\infty \left[ V_l(a) \, \partial_\theta Y_l^m - i U_l(a) \frac{\partial_\phi Y_l^m}{\sin \theta} \right] e^{i \omega t}, \\
    \xi^\phi &= \frac{1}{a^2 \sin^2 \theta} \sum_{l = |m|}^\infty \left[ V_l(a) \, \partial_\phi Y_l^m + i U_l(a) \sin \theta \, \partial_\theta Y_l^m \right] e^{i \omega t},
\end{align}
\end{subequations}
where $\omega$ is the angular frequency of the mode and $Y_l^m(\theta, \phi)$ is a spherical harmonic. In the $\Omega \rightarrow 0$ limit, the displacement vector~\eqref{eqs:displacement} will tend towards the familiar vector decomposition in $(r, \theta, \phi)$ coordinates \citep[see, e.g.,][]{1999ApJ...521..764L}, where $W_l$ and $V_l$ are the polar functions and $U_l$ is the axial function. Indeed, since a mode's parity does not change as it varies continuously along a sequence of equilibrium configurations, beginning with a spherical star and with increasing rotation, it is appropriate to identify $(W_l, V_l)$ and $U_l$ with the polar and axial perturbations, respectively, as they correspond to these classes on a spherical star.%
\footnote{This issue was somewhat confused in the critique of the formulation of \citet{1982ApJ...256..717S} by \citet{1983A&A...125..193S}.}
The scalar perturbations are consequently decomposed as 
\begin{equation}
    \delta \rho = \sum_{l = |m|}^\infty \delta \rho_l(a) Y_l^m e^{i \omega t}, \qquad \delta p = \sum_{l = |m|}^\infty \delta p_l(a) Y_l^m e^{i \omega t}, \qquad \delta \Phi = \sum_{l = |m|}^\infty \delta \Phi_l(a) Y_l^m e^{i \omega t}.
    \label{eq:scalars}
\end{equation}

\subsection{Stratification on the spherical star}

Although we want to formulate the \textit{r}-mode problem on a rotating star, stratification already plays an important role on the non-rotating star. It is instructive to consider this case briefly.

A barotropic ($\Gamma_1 = \Gamma$), spherically symmetric star only admits \textit{f}- and \textit{p}-modes. It can be shown that there are time-independent solutions to the $\Omega = 0$ perturbation equations~\eqref{eqs:perturbed} with vanishing perturbed mass densities that are purely polar or axial in nature \citep{1999ApJ...521..764L} \citep[see also][ for the corresponding result in relativity]{2000PhRvD..63b4019L}. These stationary currents are associated with the polar \textit{g}-modes and the axial \textit{r}-modes and they reside in the zero-frequency subspace on such a star. If the star spins up, these stationary currents will become oscillatory, being restored by the Coriolis force. In general, these polar and axial perturbations will mix, forming the inertial modes. However, there will persist perturbations that are purely axial at zeroth order in rotation, associated with a definite $(l, m)$. These solutions are the fundamental $l = |m|$ \textit{r}-modes.

Suppose we now consider a stratified ($\Gamma_1 \neq \Gamma$), spherical star. Alongside the \textit{f}- and \textit{p}-modes, it will host \textit{g}-modes with non-zero frequencies, supported by the buoyancy. The only trivial solutions that exist in this case are axial. Therefore, when the star rotates, the only inertial modes that appear are the \textit{r}-modes. Because the axial perturbations do not have stationary polar currents to mix with, a non-barotropic, rotating star has the complete set of $l \geq |m|$ \textit{r}-modes including radial overtones.

In between these two extremal cases, there is a third regime where the stratification and rotation are of the same order of magnitude. This is a form of weak (but non-zero) stratification relative to the rotation. We discuss this scenario in more detail in \citet{other}.

\begin{figure}
    \centering
    \includegraphics[width=0.5\columnwidth]{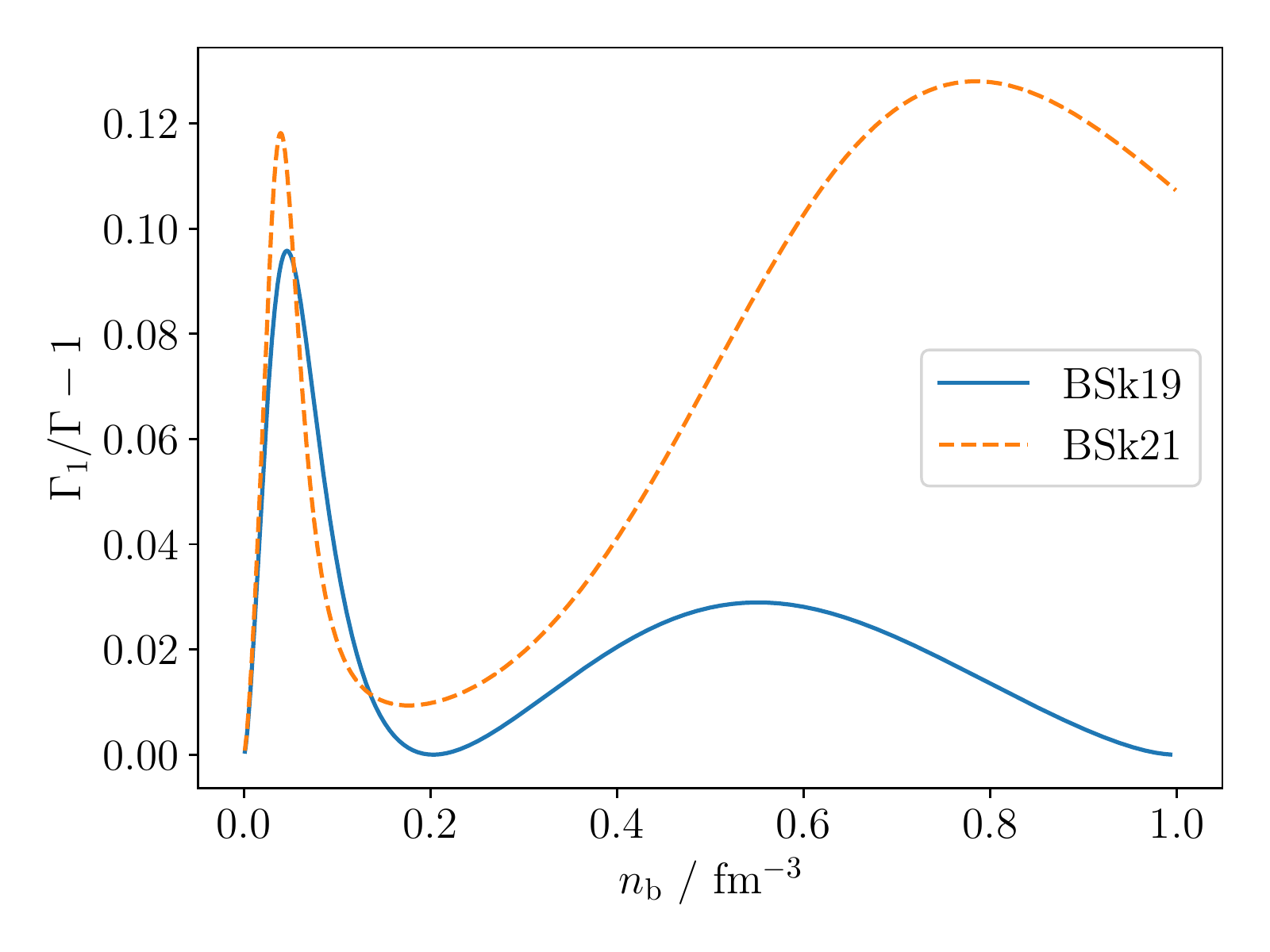}
    \caption{\label{fig:frac}%
    The relative difference between the adiabatic indices of the equilibrium $\Gamma$ and the perturbations $\Gamma_1$ against the baryon-number density $n_\text{b}$ for two nuclear-matter equations of state, BSk19 and BSk21. These models show that, although neutron stars are predominantly non-barotropic, they are expected to have barotropic layers. In particular, both equations of state are barotropic towards the surface and BSk19 is also close to barotropic near $n_\text{b} = 0.2 \ \text{fm}^{-3}, 1 \ \text{fm}^{-3}$. See \citet{other} for a description of how $\Gamma_1$ is defined.}
\end{figure}

Neutron stars are, in general, non-barotropic since the chemical composition changes throughout the interior. This is illustrated in Fig.~\ref{fig:frac} for the two realistic equations of state, BSk19 and BSk21 \citep{2013A&A...559A.128F,2013A&A...560A..48P}. The exact chemical composition of neutron-star interiors is at present unknown and is related to the nuclear reactions going on under the surface \citep{1992ApJ...395..240R,2019MNRAS.489.4043A}. However, we understand low-density nuclear matter quite well and expect the outer layers to be barotropic.(This is shown in Fig.~\ref{fig:frac} for both equations of state where $\Gamma_1 = \Gamma$ at low densities.)

\subsection{Slow rotation}

Moving beyond the spherical star, in a slow-rotation expansion, an \textit{r}-mode has the ordering%
\footnote{It is straightforward to show from a slow-rotation expansion of the linearised equations~\eqref{eqs:perturbed} that an axial perturbation at zeroth order in rotation with a frequency at linear order will source polar perturbations at second order.}
\begin{equation}
\begin{gathered}
    \omega \sim O(\epsilon), \qquad W_{l'} \sim O(\epsilon^2), \qquad V_{l'} \sim O(\epsilon^2), \qquad U_{l'} \sim O(1), \\
    \delta \rho_{l'} \sim O(\epsilon^2), \qquad \delta p_{l'} \sim O(\epsilon^2), \qquad \delta \Phi_{l'} \sim O(\epsilon^2).
\end{gathered}
    \label{eq:ordering}
\end{equation}
To uniquely identify the \textit{r}-modes from the perturbation equations~\eqref{eqs:perturbed}, it is sufficient to look for solutions of the form~\eqref{eq:ordering}. By rotational symmetry, we can assume the series expansion for the mode frequency \citep{1978MNRAS.182..423P,1981Ap&SS..78..483S} 
\begin{equation}
    \omega = \omega_0 + \omega_2 + O(\epsilon^5),
    \label{eq:frequencyExpansion}
\end{equation}
where $\omega_0 = O(\epsilon)$ and $\omega_2 = O(\epsilon^3)$. Clearly, the validity of this expansion relies on $|\omega_2 / \omega_0| = O(\epsilon^2) \ll 1$. (This is a feature we will pay close attention to later.) The remaining terms we need only calculate to leading order.

In addition to the fact that the \textit{r}-modes are purely axial, they are special among the inertial modes in that their frequency at leading order is analytic and independent of the equation of state. This can be seen from the angular components of the linearised Euler equation~\eqref{eq:perturbedEuler}. At $O(\epsilon^2)$, we find the simple expression
\begin{equation}
    0 = \omega_0 \sum_l [l (l + 1) \omega_0 - 2 m \Omega] U_l Y_l^m.
    \label{eq:curl1}
\end{equation}
There are three solutions to equation~\eqref{eq:curl1}: (i) the equation permits a zero-frequency solution with $\omega_0 = 0$; (ii) the axial functions are $U_{l'} = 0$ for all $l'$; or (iii) a single $U_l$ survives with frequency~\eqref{eq:omega_0}. The physically interesting case that corresponds to an \textit{r}-mode is solution~(iii). We note that there are no axisymmetric ($m = 0$) \textit{r}-modes and all the perturbations travel retrograde to the rotation of the star. However, at this order, the axial function $U_l$ is undetermined. To calculate these functions, we need to develop the perturbation equations~\eqref{eqs:perturbed} into an eigenvalue problem for the frequency correction $\omega_2$.

The equations we need were first derived by \citet{1982ApJ...256..717S}. We will focus on the simple adiabatic case appropriate for (cold) neutron stars. In this context, the oscillations will be normal modes with manifestly real frequencies, since there are no dissipative effects. As is typically useful in the numerical computation of oscillation modes, we define the following dimensionless variables \citep{1989nos..book.....U}:
\begin{equation}
    y_{1, l'} = \frac{W_{l'}}{a^2}, \qquad y_{2, l'} = \frac{1}{a g} \bigg( \frac{\delta p_{l'}}{\rho_0} + \delta \Phi_{l'} \bigg), \qquad y_{3, l'} = \frac{1}{a g} \delta \Phi_{l'}, \qquad y_{4, l'} = \frac{1}{g} \frac{d \delta \Phi_{l'}}{da}, \qquad y_{5, l'} = \frac{V_{l'}}{a^2}, \qquad y_{6, l'} = \frac{U_{l'}}{a^2},
\end{equation}
where $g = d \Phi_0 / da$ is the gravitational acceleration. Rotation couples spherical harmonics with different values of $l'$. In order to see this, one may use the standard recurrence relations
\begin{subequations}\label{eqs:recursion}
\begin{align}
    \cos \theta \, Y_{l'}^m &= Q_{l' + 1} Y_{l' + 1}^m + Q_{l'} Y_{l' - 1}^m, \\
    \sin \theta \, \partial_\theta Y_{l'}^m &= l' Q_{l' + 1} Y_{l' + 1}^m - (l' + 1) Q_{l'} Y_{l' - 1}^m,
\end{align}
\end{subequations}
where 
\begin{equation}
    Q_{l'} = \sqrt{\frac{(l' + m) (l' - m)}{(2 l' + 1) (2 l' - 1)}}.
\end{equation}
It is useful to note the orthogonality of the spherical harmonics 
\begin{equation}
    \int Y_{l'}^{m'} Y_l^{m *} d\Omega = \delta_{l l'} \delta_{m m'},
    \label{eq:orthogonality}
\end{equation}
where $d\Omega$ is the element of solid angle. Using these identities, the linearised equations~\eqref{eqs:perturbed} with the ordering~\eqref{eq:ordering} provide the following system of differential equations:
\begin{subequations}\label{eqs:r-mode}
\begin{align}
    a \frac{d y_{1, l + 1}}{da} &=  \left( \frac{V}{\Gamma_1} - 3 \right) y_{1, l + 1} - \frac{V}{\Gamma_1} y_{2, l + 1} + \frac{V}{\Gamma_1} y_{3, l + 1} + (l + 1) (l + 2) y_{5, l + 1} + \frac{3 m}{a^2} \frac{d (a^3 \varepsilon_2)}{da} Q_{l + 1} y_{6, l}, \label{eq:continuityPlus}\\
    a \frac{d y_{1, l - 1}}{da} &= \left( \frac{V}{\Gamma_1} - 3 \right) y_{1, l - 1} - \frac{V}{\Gamma_1} y_{2, l - 1} + \frac{V}{\Gamma_1} y_{3, l - 1} + (l - 1) l y_{5, l - 1} + \frac{3 m}{a^2} \frac{d (a^3 \varepsilon_2)}{da} Q_l y_{6, l}, \label{eq:continuityMinus}\\
    a \frac{d y_{2, l + 1}}{da} &= a A y_{1, l + 1} + (1 - U - a A) y_{2, l + 1} + a A y_{3, l + 1} - 2 l Q_{l + 1} c_1 \epsilon \tilde{\omega}_0 y_{6, l}, \label{eq:Euler-aPlus}\\
    a \frac{d y_{2, l - 1}}{da} &= a A y_{1, l - 1} + (1 - U - a A) y_{2, l - 1} + a A y_{3, l - 1} + 2 (l + 1) Q_l c_1 \epsilon \tilde{\omega}_0 y_{6, l}, \label{eq:Euler-aMinus}\\
    a \frac{d y_{3, l \pm 1}}{da} &= (1 - U) y_{3, l \pm 1} + y_{4, l \pm 1}, \\
    a \frac{d y_{4, l \pm 1}}{da} &= - a A U y_{1, l \pm 1} + \frac{U V}{\Gamma_1} y_{2, l \pm 1} + \left[ (l \pm 1) (l \pm 1 + 1) - \frac{U V}{\Gamma_1} \right] y_{3, l \pm 1} - U y_{4, l \pm 1},
\end{align}
with the algebraic relations 
\begin{gather}
    (l + 1) y_{2, l + 1} = - 2 l Q_{l + 1} c_1 \epsilon \tilde{\omega}_0 y_{6, l}, \label{eq:divergencePlus}\\
    l y_{2, l - 1} = - 2 (l + 1) Q_l c_1 \epsilon \tilde{\omega}_0 y_{6, l}, \label{eq:divergenceMinus}\\
    B y_{6, l} = - (l + 1) Q_l [y_{1, l - 1} - (l - 1) y_{5, l - 1}] + l Q_{l + 1} [y_{1, l + 1} + (l + 2) y_{5, l + 1}], \label{eq:curl2}
\end{gather}
\end{subequations}
where 
\begin{equation}
    A = \frac{d \ln \rho_0}{da} - \frac{1}{\Gamma_1} \frac{d \ln p_0}{da} = \left( \frac{1}{\Gamma} - \frac{1}{\Gamma_1} \right) \frac{d \ln p_0}{da}
\end{equation}
is the Schwarzschild discriminant (useful for characterising the stratification of the fluid) and we have defined the dimensionless terms
\begin{equation}
\begin{split}
    &\tilde{\omega}_0 = \frac{\omega_0}{\sqrt{G M / R^3}}, \qquad \tilde{\omega}_2 = \frac{\omega_2}{\sqrt{G M / R^3}}, \\
    &V = - \frac{d \ln p_0}{d \ln a}, \qquad U = \frac{d \ln m_0}{d \ln a}, \qquad c_1 = \left( \frac{a}{R} \right)^3 \frac{M}{m_0}, \\
    &B = \frac{1}{2 \epsilon \tilde{\omega}_0} \{ 2 [l (l + 1) \tilde{\omega}_0 - m \epsilon] \tilde{\omega}_2 - [l (l + 1) \tilde{\omega}_0 - 8 m \epsilon] \tilde{\omega}_0 \varepsilon_2 + 3 (Q_{l + 1}^2 + Q_l^2) [l (l + 1) \tilde{\omega}_0 - 12 m \epsilon] \tilde{\omega}_0 \varepsilon_2 + 6 [l Q_{l + 1}^2 - (l + 1) Q_l^2] \tilde{\omega}_0^2 \varepsilon_2 \}.
\end{split}
\end{equation}
Equations~\eqref{eqs:r-mode} describe an \textit{r}-mode and there are a few remarks worth making: (i) the leading-order axial perturbation, associated with the spherical harmonic $Y_l^m$, sources $l \pm 1$ polar perturbations at $O(\epsilon^2)$ (as well as $l \pm 2$ axial perturbations at this order, as we show in Appendix~\ref{app:axial}); (ii) since $Q_l = 0$ for $l = |m|$, the $l - 1$ polar terms must vanish and thus the lowest degree mode that exists is $l = |m|$ [this has been assumed in the decomposition of equations~\eqref{eqs:displacement} and \eqref{eq:scalars}]; and (iii) the only rotational quantity from the background that comes into the mode equations is the non-spherical shape $\varepsilon_2$. At this point, no assumptions have been made about the state of the matter.

In the globally barotropic case $\Gamma_1 = \Gamma$, where $A = 0$, the linearised equations~\eqref{eqs:r-mode} admit a simple solution. We can combine \eqref{eq:Euler-aPlus} with \eqref{eq:divergencePlus} and \eqref{eq:Euler-aMinus} with \eqref{eq:divergenceMinus} to obtain two equations that an \textit{r}-mode must satisfy,
\begin{subequations}\label{eqs:barotrope}
\begin{align}
    Q_{l + 1} \left[ a \frac{d y_{6, l}}{da} - (l - 1) y_{6, l} \right] &= 0, \\ 
    Q_l \left[ a \frac{d y_{6, l}}{da} + (l + 2) y_{6, l} \right] &= 0.
\end{align}
\end{subequations}
Equations~\eqref{eqs:barotrope} only provide a consistent, non-trivial solution when $l = |m|$ and $y_{6, |m|} \propto a^{|m| - 1}$. This is the fundamental \textit{r}-mode that we alluded to previously. Thus, there are no overtones when the star is barotropic and there are no $l > |m|$ solutions. The higher order features, including the other eigenfunctions and eigenvalue $\tilde{\omega}_2$, are determined from the remaining $l + 1$ equations~\eqref{eqs:r-mode}.

\section{Globally non-barotropic stars}
\label{sec:non-barotropic}

We want to express the \textit{r}-mode equations~\eqref{eqs:r-mode} in a form that is suitable for integration. As it stands, we are unable to determine $y_{5, l \pm 1}$, since we only have one algebraic relation~\eqref{eq:curl2} for these two functions that appear in equations~\eqref{eq:continuityPlus} and \eqref{eq:continuityMinus}. To begin with, we will assume the star to be stably stratified throughout its interior such that $\Gamma_1 \neq \Gamma$.%
\footnote{This is equivalent to assuming $A \neq 0$, except at the very centre of the star. However, due to the coordinate system, the equations are divergent at the centre, so one usually circumvents this singularity in the numerical integration by considering a small step away from $a = 0$.}
As we have noted above, this assumption is inappropriate for realistic stars with some degree of local barotropicity, but we can learn something about how the \textit{r}-mode solutions behave as one gets close to the barotropic limit.

The variables $y_{1, l - 1}$, $y_{2, l - 1}$, $y_{5, l \pm 1}$ and $y_{6, l}$ can be eliminated from the system of equations through the following approach \citep{1982ApJ...256..717S}. Starting with equations~\eqref{eq:divergencePlus} and \eqref{eq:divergenceMinus}, we have an algebraic relation for $y_{6, l}$ and 
\begin{equation}
    y_{2, l - 1} = \mathcal{J}_l y_{2, l + 1},
\end{equation}
where 
\begin{equation}
    \mathcal{J}_l = \frac{(l + 1)^2}{l^2} \frac{Q_l}{Q_{l + 1}}.
\end{equation}
We can make use of these relations with equations~\eqref{eq:Euler-aPlus} and \eqref{eq:Euler-aMinus} in order to obtain 
\begin{equation}
    a A y_{1, l - 1} = \mathcal{J}_l a A y_{1, l + 1} + \mathcal{J}_l (2 l + 1) y_{2, l + 1} + a A (\mathcal{J}_l y_{3, l + 1} - y_{3, l - 1}).
    \label{eq:relation-y1lminus1}
\end{equation}
Since we assume that $A \neq 0$, this allows us to eliminate $y_{1, l - 1}$ from the system. Here, we note that \eqref{eq:relation-y1lminus1} implies that $y_{2, l + 1} = 0$ in barotropic regions, for $l > |m|$. Next, we can differentiate \eqref{eq:relation-y1lminus1} and combine the result with the differential equations~\eqref{eq:continuityPlus} and \eqref{eq:continuityMinus} to obtain 
\begin{equation}
\begin{split}
    (l - 1) l y_{5, l - 1} = \mathcal{J}_l& (2 l + 1) y_{1, l + 1} + \mathcal{J}_l (5 + 2 l - U) y_{3, l + 1} - (4 - U) y_{3, l - 1} + \mathcal{J}_l y_{4, l + 1} - y_{4, l - 1} + \mathcal{J}_l (l + 1) (l + 2) y_{5, l + 1} \\
    &+ \mathcal{J}_l (2 l + 1) \bigg\{ \frac{1}{a A} \left[ 5 + l - U - \frac{V}{\Gamma_1} - a A - \frac{1}{A} \frac{d (a A)}{da} \right] - \frac{3 m}{2 l (l + 1) a^2 c_1 \epsilon \tilde{\omega}_0} \frac{d (a^3 \varepsilon_2)}{da} \bigg\} y_{2, l + 1}.
\end{split}
\end{equation}
Along with equation~\eqref{eq:curl2}, we are in a position to decouple $y_{5, l \pm 1}$.

We now have enough information to remove the variables $y_{1, l - 1}$, $y_{2, l - 1}$, $y_{5, l \pm 1}$ and $y_{6, l}$ from equations~\eqref{eqs:r-mode}. We end up with
\begin{subequations}\label{eqs:r-mode-non-barotropic}
\begin{align}
\begin{split}
    a \frac{d y_{1, l + 1}}{da} &= \left( \frac{V}{\Gamma_1} - 4 - l \right) y_{1, l + 1} - \left[ \frac{(2 l + 1) C}{1 + \mathcal{J}_l^2} + \frac{V}{\Gamma_1} + \frac{3 m (l + 1)}{2 l a^2 c_1 \epsilon \tilde{\omega}_0} \frac{d (a^3 \varepsilon_2)}{da} \right] y_{2, l + 1} \\
    &\quad+ \left[ \frac{V}{\Gamma_1} + \frac{\mathcal{J}_l^2}{1 + \mathcal{J}_l^2} (U - 5 - l) \right] y_{3, l + 1}
    + \frac{\mathcal{J}_l}{1 + \mathcal{J}_l^2} (4 - U - l) y_{3, l - 1} - \frac{\mathcal{J}_l^2}{1 + \mathcal{J}_l^2} y_{4, l + 1} + \frac{\mathcal{J}_l}{1 + \mathcal{J}_l^2} y_{4, l - 1},
\end{split}\\
    a \frac{d y_{2, l + 1}}{da} &= a A y_{1, l + 1} + (2 + l - U - a A) y_{2, l + 1} + a A y_{3, l + 1}, \\
    a \frac{d y_{3, l \pm 1}}{da} &= (1 - U) y_{3, l \pm 1} + y_{4, l \pm 1}, \\
    a \frac{d y_{4, l + 1}}{da} &= - a A U y_{1, l + 1} + \frac{U V}{\Gamma_1} y_{2, l + 1} + \left[ (l + 1) (l + 2) - \frac{U V}{\Gamma_1} \right] y_{3, l + 1} - U y_{4, l + 1}, \\
    a \frac{d y_{4, l - 1}}{da} &= - \mathcal{J}_l a A U y_{1, l + 1} + \mathcal{J}_l U \left( \frac{V}{\Gamma_1} - 2 l - 1 \right) y_{2, l + 1} - \mathcal{J}_l a A U y_{3, l + 1} + \left[ (l - 1) l - \frac{U V}{\Gamma_1} + a A U \right] y_{3, l - 1} - U y_{4, l - 1},
\end{align}
\end{subequations}
where 
\begin{equation}
    C = \frac{\mathcal{J}_l^2}{a A} \left[ 5 - U - \frac{V}{\Gamma_1} - \frac{1}{A} \frac{d (a A)}{da} \right] - \mathcal{J}_l^2 + \frac{1}{2 l c_1 \epsilon \tilde{\omega}_0} \left[ - \frac{3 m \mathcal{J}_l^2}{(l + 1) a^2} \frac{d (a^3 \varepsilon_2)}{da} + \frac{(l + 1)^2 B}{l (2 l + 1) Q_{l + 1}^2} \right].
\end{equation}

In order to solve the eigenvalue problem~\eqref{eqs:r-mode-non-barotropic}, we must provide boundary conditions that constrain the solutions. At the stellar centre, the functions must be well behaved and regular. As $a \rightarrow 0$, the regular solutions are given by \citep[see Section~18.1 of][]{1989nos..book.....U}
\begin{equation}
    y_{1, l + 1} \sim a^{l - 1}, \qquad y_{2, l + 1} \sim a^{l - 1}, \qquad y_{3, l \pm 1} \sim a^{l \pm 1 - 2}, \qquad y_{4, l \pm 1} \sim a^{l \pm 1 - 2}.
    \label{eq:centreFunctions}
\end{equation}
Inserting \eqref{eq:centreFunctions} into the perturbation equations~\eqref{eqs:r-mode-non-barotropic}, we find the following boundary conditions at the centre of the star:
\begin{subequations}\label{eqs:centre}
\begin{gather}
    (2 l + 3) y_{1, l + 1} + \left[ \frac{(2 l + 1) C}{1 + \mathcal{J}_l^2} + \frac{9 m (l + 1) \varepsilon_2}{2 l c_1 \epsilon \tilde{\omega}_0} \right] y_{2, l + 1} + \frac{\mathcal{J}_l^2}{1 + \mathcal{J}_l^2} (2 l + 3) y_{3, l + 1} = 0, \\
    (l \pm 1) y_{3, l \pm 1} - y_{4, l \pm 1} = 0.
\end{gather}
\end{subequations}
At the surface, the Lagrangian variation of the pressure must vanish $\Delta p / \rho_0 = 0$ and the perturbed gravitational potential must match smoothly to the exterior solution, which decays as $\delta \Phi_{l'} \propto 1 / r^{l' + 1}$ for a given multipole $l'$. Thus, for $a = R$, we must have
\begin{subequations}\label{eqs:surface}
\begin{align}
    y_{1, l + 1} - y_{2, l + 1} + y_{3, l + 1} &= 0, \\
    (l \pm 1 + 1) y_{3, l \pm 1} + y_{4, l \pm 1} &= 0.
\end{align}
\end{subequations}
Equations~\eqref{eqs:r-mode-non-barotropic} constitute six first-order ordinary differential equations, supplemented by three boundary conditions at the centre~\eqref{eqs:centre} and three at the surface~\eqref{eqs:surface}. The boundary conditions will only be satisfied for an eigenvalue $\tilde{\omega}_2$.

We solve this eigenvalue problem using the following numerical approach \citep[our technique is similar to that used by][]{1983ApJS...53...73L}. Since the system of equations~\eqref{eqs:r-mode-non-barotropic} is linear (by construction, as we are using first-order perturbation theory), we may express it as
\begin{equation}
    \frac{d\mathbf{Y}}{da} = \mathbf{Q}(a; l, m, \tilde{\omega}_2) \cdot \mathbf{Y}(a),
    \label{eq:linearSystem}
\end{equation}
with a matrix $\mathbf{Q}$ and abstract vector field $\mathbf{Y} = (y_{1, l + 1}, y_{2, l + 1}, y_{3, l + 1}, y_{3, l - 1}, y_{4, l + 1}, y_{4, l - 1})$. This is a sixth-order system of linear equations, so there will exist six linearly independent solutions for a given $(l, m, \tilde{\omega}_2)$. However, only for specific values of $\tilde{\omega}_2$ will the linearly independent solutions combine to satisfy all the boundary conditions; these are the eigenfrequencies. At the centre of the star, there are three linearly independent vectors $\mathbf{Y}(a \rightarrow 0)$ that satisfy boundary conditions~\eqref{eqs:centre}. We select three such initial vectors and integrate them using equation~\eqref{eq:linearSystem} to a point $0 < a_0 < R$ in the star. This generates three solutions $\mathbf{Y}_1$, $\mathbf{Y}_2$, $\mathbf{Y}_3$ defined on the domain $0 < a \leq a_0$, each of which satisfy the central boundary conditions~\eqref{eqs:centre}. In a similar fashion, we also produce three linearly independent solutions $\mathbf{Y}_4$, $\mathbf{Y}_5$, $\mathbf{Y}_6$ for the region $a_0 \leq a \leq R$ out of the surface boundary conditions~\eqref{eqs:surface}. Therefore, we obtain the general solution
\begin{equation}
    \mathbf{Y}(a) = 
    \begin{cases}
        \beta_1 \mathbf{Y}_1(a) + \beta_2 \mathbf{Y}_2(a) + \beta_3 \mathbf{Y}_3(a), & \text{for } 0 < a \leq a_0 \\
        \beta_4 \mathbf{Y}_4(a) + \beta_5 \mathbf{Y}_5(a) + \beta_6 \mathbf{Y}_6(a), & \text{for } a_0 \leq a \leq R
    \end{cases}
\end{equation}
with constants $\beta_1, \ldots, \beta_6$. Hence, $\tilde{\omega}_2$ is an eigenvalue if and only if
\begin{equation}
    \beta_1 \mathbf{Y}_1(a_0) + \beta_2 \mathbf{Y}_2(a_0) + \beta_3 \mathbf{Y}_3(a_0) = \beta_4 \mathbf{Y}_4(a_0) + \beta_5 \mathbf{Y}_5(a_0) + \beta_6 \mathbf{Y}_6(a_0).
\end{equation}
This matching can be written as the matrix equation $\mathbf{A} \cdot \mathbf{x} = 0$, where $\mathbf{A}$ depends non-linearly on $\tilde{\omega}_2$. For non-trivial eigenvalues, the matrix $\mathbf{A}$ is singular. Thus, we look for values of $\tilde{\omega}_2$, for a given $(l, m)$, such that $\det{\mathbf{A}} = 0$.

Once the eigenfrequency $\tilde{\omega}_2$ is determined, the eigenfunctions may be calculated. Since we are calculating normal modes, there is a free amplitude in the functions. We choose to normalise the modes at the surface by
\begin{equation}
    y_{6, l}(R) = 1.
\end{equation}
We implement this normalisation by replacing a row in $\mathbf{A}$ in favour of this condition, $\mathbf{\tilde{A}} \cdot \mathbf{x} = \mathbf{b}$, where $\mathbf{\tilde{A}}$ becomes a non-singular matrix and $\mathbf{b}$ is a known column vector. This concludes the discussion of our numerical method.

For our results, we assume a polytropic equation of state for the equilibrium star
\begin{equation}
    p(\rho) = K \rho^\Gamma, \qquad \Gamma = 1 + \frac{1}{n},
\end{equation}
where $K$ and $n$ are the polytropic constant and index, respectively. We consider $n = 1$ to approximate a neutron star and obtain the shape $\varepsilon_2$ from the results of \citet{1962ApJ...136.1082C}.

As a consistency check of our computational technique, we first consider $\Gamma_1 = 5/3$ \citep[also employed by][]{1981A&A....94..126P,1982ApJ...256..717S}.%
\footnote{Such a star is unstable to convective phenomena $A > 0$. The buoyancy forces will tend to increase the displacement of fluid elements, giving rise to unstable \textit{g}-modes.} 
The results are summarised by Table~\ref{tab:Gamma153}, where the overtones, with nodes in their displacement eigenfunctions, are denoted by $k$. We can compare the $l = 3$ eigenfrequencies with \citet{1981A&A....94..126P}, who calculated within the Cowling approximation. Our results are compatible within the estimated errors of the approximation. Additionally, we considered the non-analytic $n = 3$ polytrope, numerically solving \eqref{eq:shape} for the shape, to compare with \citet{1982ApJ...256..717S}, finding excellent agreement.

\begin{table}
\centering
\caption{The \textit{r}-mode eigenfrequencies $\tilde{\omega}_2 / \epsilon^3$ of the $n = 1$ polytrope with $\Gamma_1 = 5/3$.}
\label{tab:Gamma153}
\begin{tabular}{ c r r r r r r }
\hline
 & \multicolumn{3}{c}{$l = 3$} & \multicolumn{2}{c}{$l = 2$} & \multicolumn{1}{c}{$l = 1$} \\ \cline{2-4} \cline{5-6} \cline{7-7}
$k$ & \multicolumn{1}{c}{$m = 3$} & \multicolumn{1}{c}{$m = 2$} & \multicolumn{1}{c}{$m = 1$} & \multicolumn{1}{c}{$m = 2$} & \multicolumn{1}{c}{$m = 1$} & \multicolumn{1}{c}{$m = 1$} \\
\hline
0 & 0.42272 & 0.98626 & 0.70223 & 0.39651 & 1.86008 & 0.00000 \\
1 & 0.69283 & 1.96099 & 1.43025 & 1.02241 & 4.36269 & 1.84339 \\
2 & 1.02524 & 3.23340 & 2.38189 & 1.81237 & 7.78243 & 4.24366 \\
3 & 1.43370 & 4.80556 & 3.55793 & 2.81025 & 12.12275 & 7.39318 \\
\hline
\end{tabular}
\end{table}

\begin{figure}
    \centering
\subfloat[$m = 3$]{%
    \includegraphics[width=0.5\columnwidth]{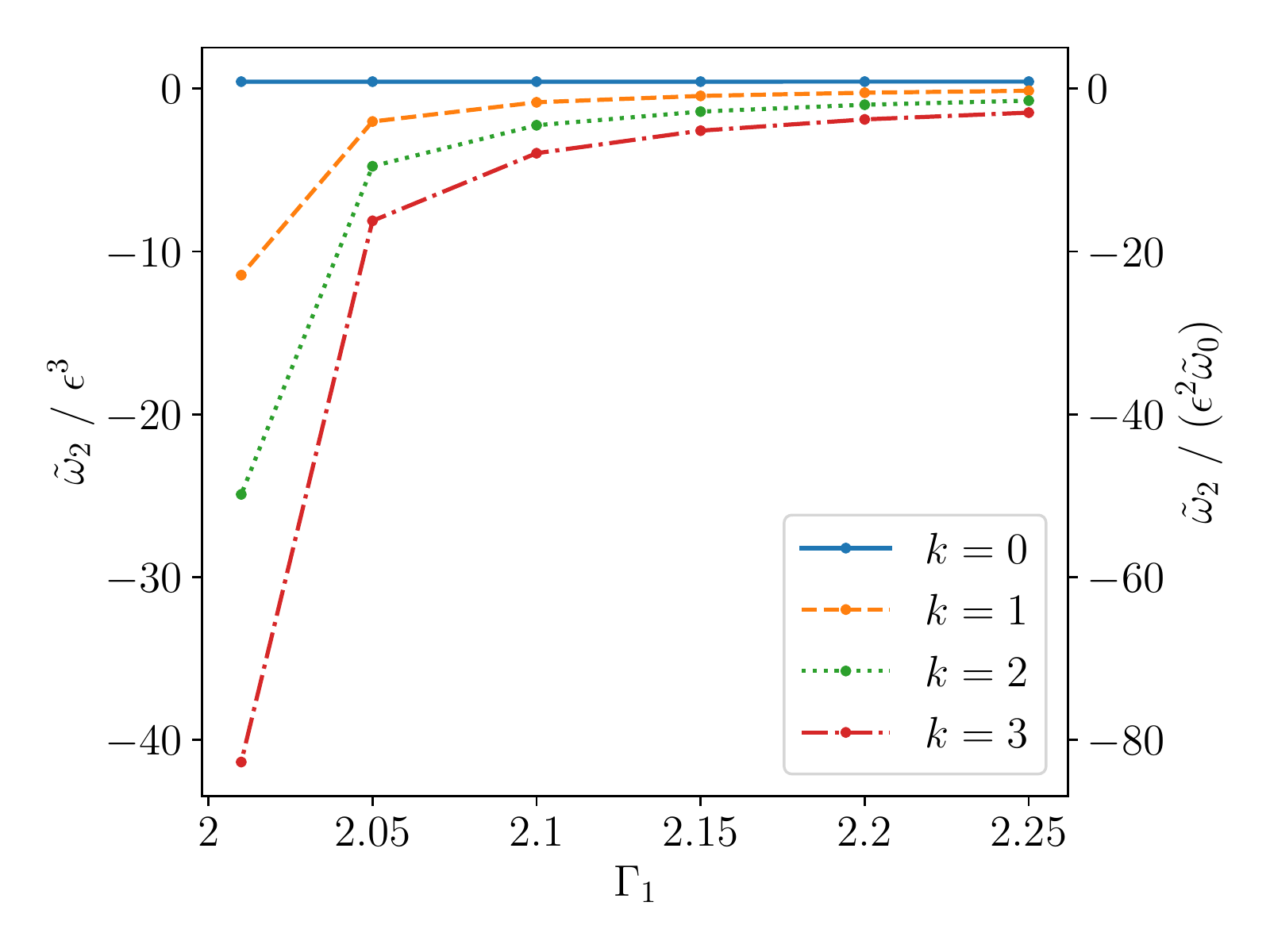}%
}
\subfloat[$m = 2$]{%
    \includegraphics[width=0.5\columnwidth]{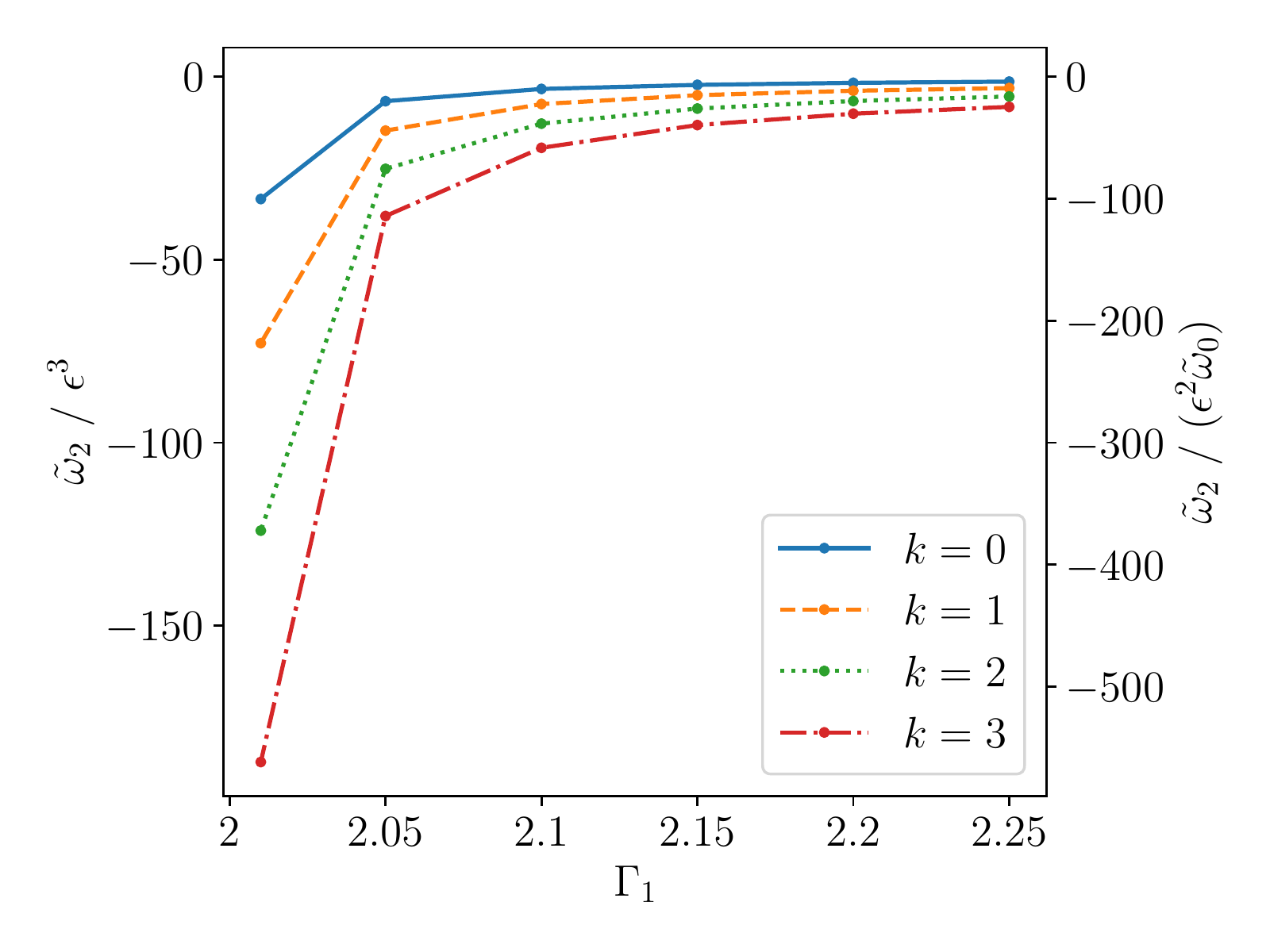}%
}\\
\subfloat[$m = 1$]{%
    \includegraphics[width=0.5\columnwidth]{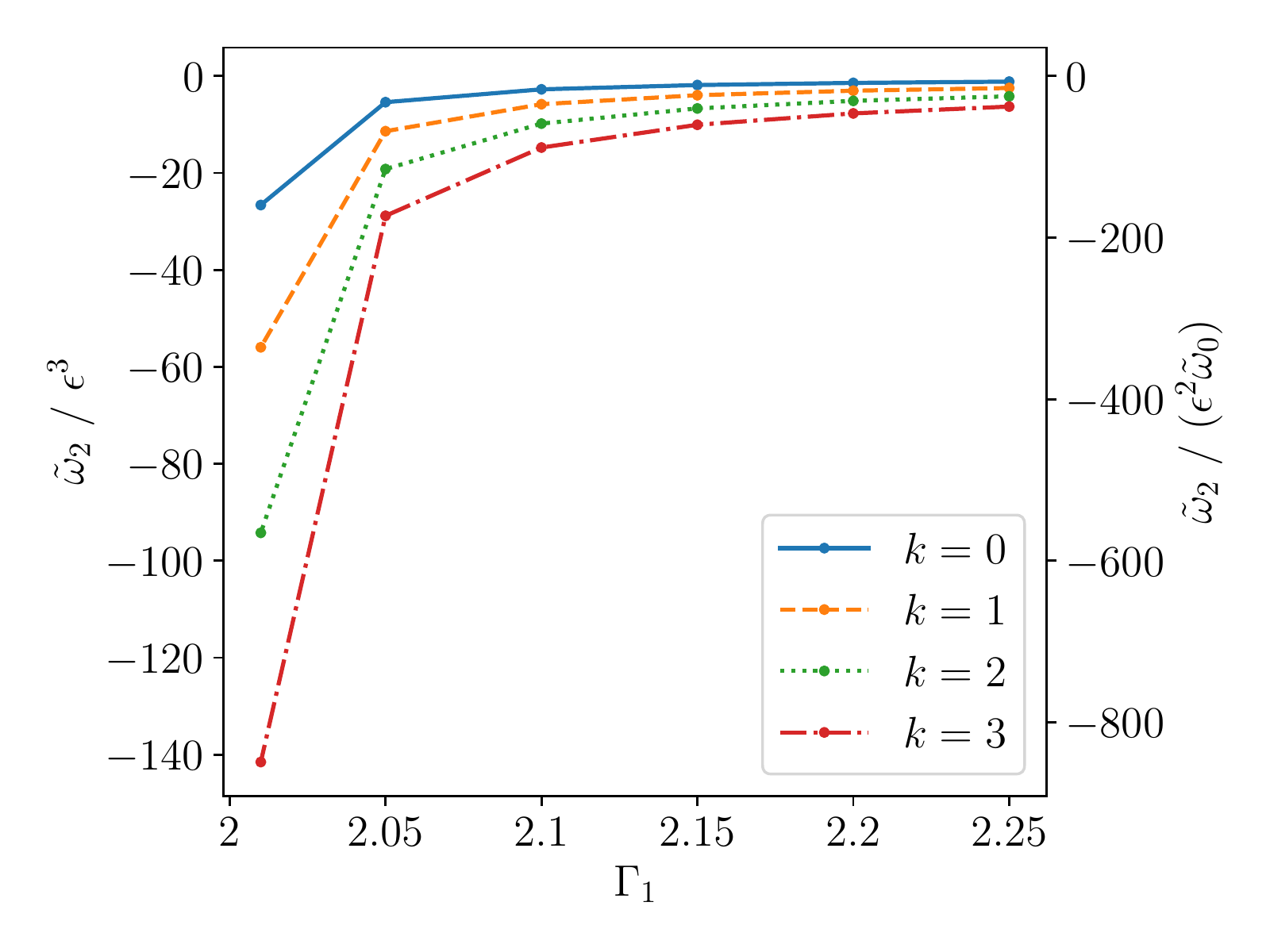}%
}
    \caption{The eigenfrequencies $\tilde{\omega}_2$ of the $l = 3$ \textit{r}-modes of the $n = 1$ polytrope with varying $\Gamma_1$. Note that only the $(m, k) = (3, 0)$ solution remains well behaved in the barotropic limit. The other solutions diverge, implying that $\tilde{\omega}_2$ is promoted to the same order as the leading-order frequency $\tilde{\omega}_0$ and the frequency expansion is spoiled.}
    \label{fig:l3}
\end{figure}

\begin{figure}
\subfloat[$m = 2$]{%
    \includegraphics[width=0.5\columnwidth]{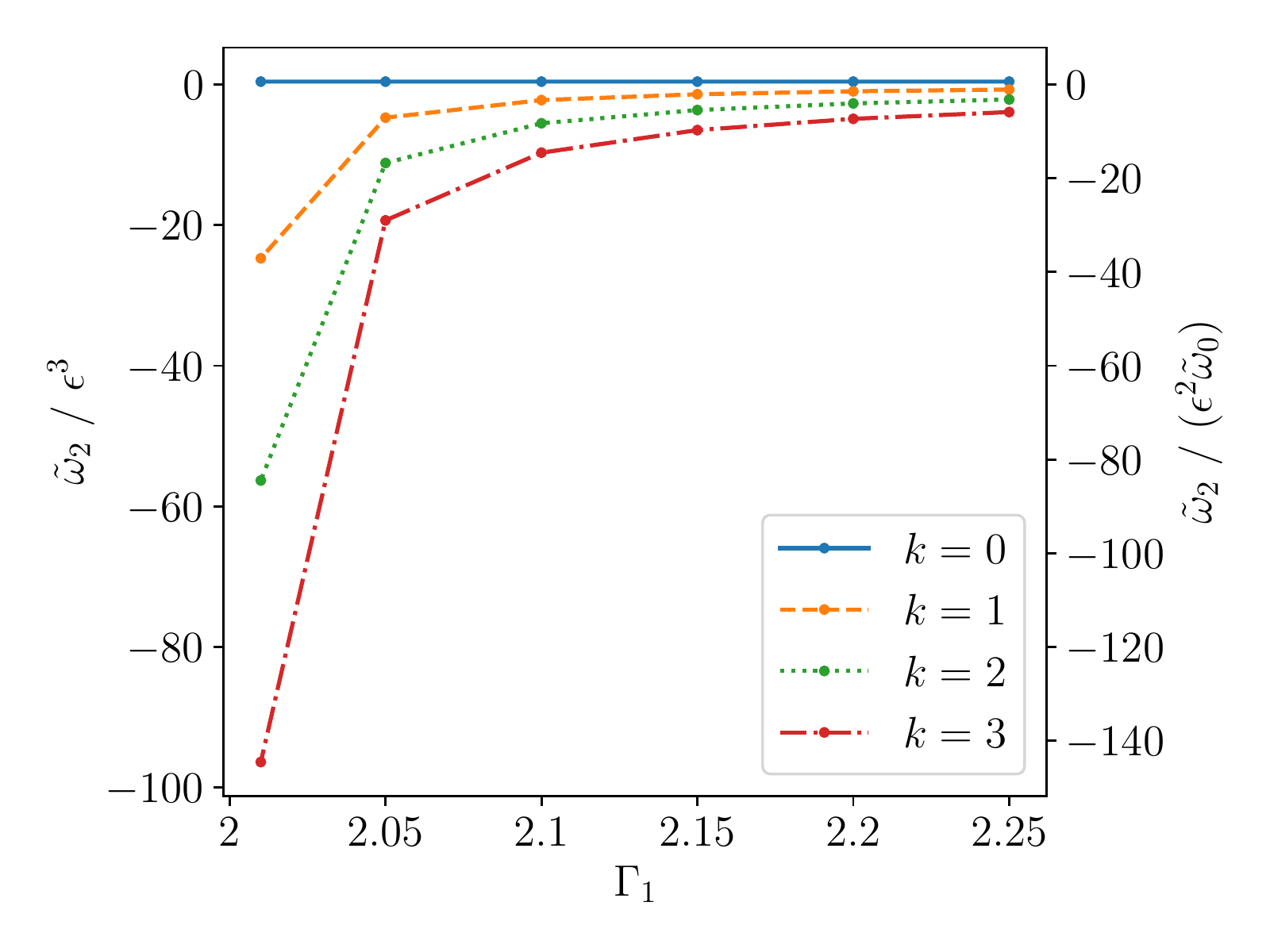}%
}
\subfloat[$m = 1$]{%
    \includegraphics[width=0.5\columnwidth]{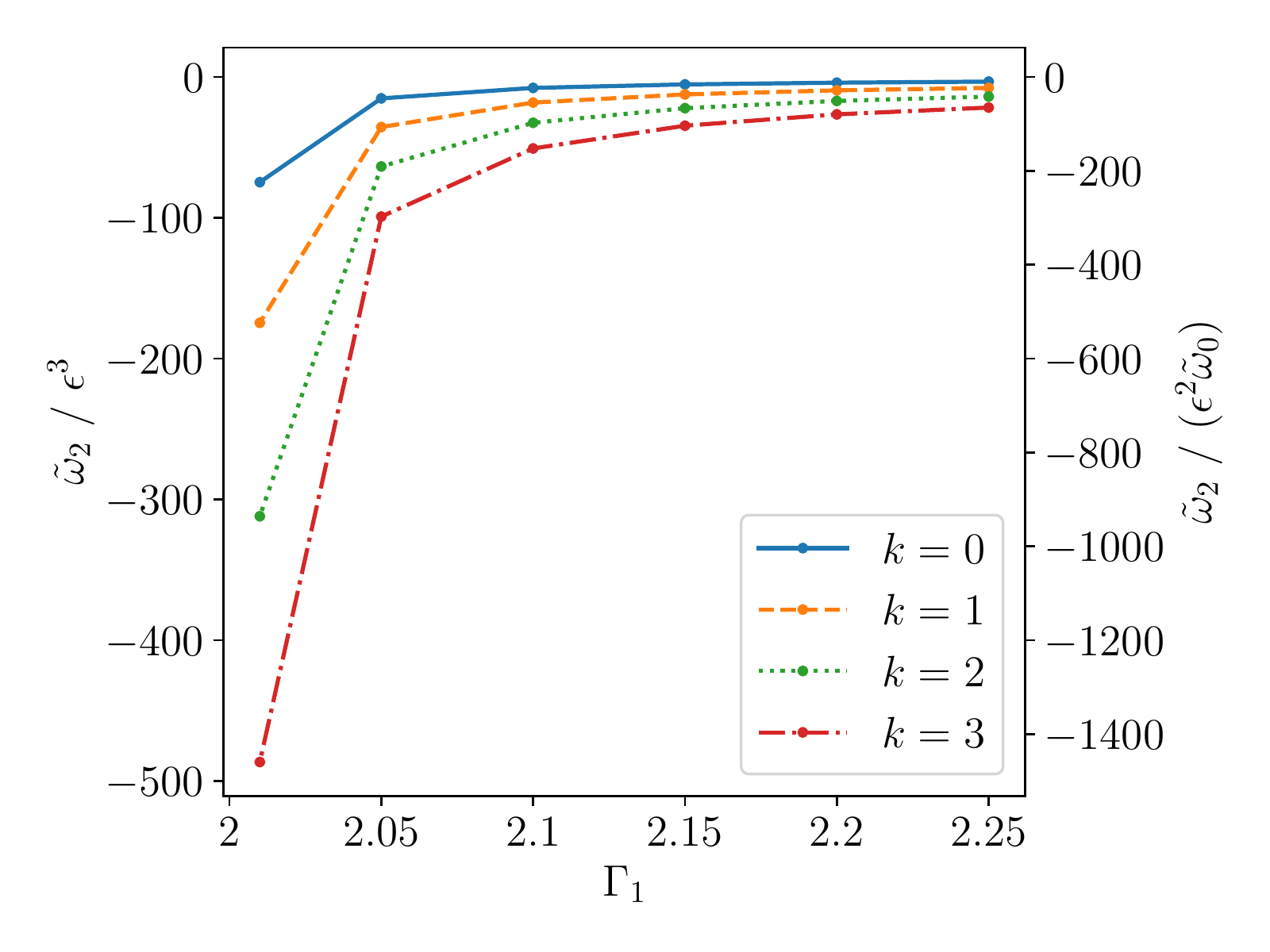}%
}
    \caption{The eigenfrequencies $\tilde{\omega}_2$ of the $l = 2$ \textit{r}-modes of the $n = 1$ polytrope with varying $\Gamma_1$. Note that only the $(m, k) = (2, 0)$ solution remains well behaved in the barotropic limit. The other solutions diverge, implying that $\tilde{\omega}_2$ is promoted to the same order as the leading-order frequency $\tilde{\omega}_0$ and the frequency expansion is spoiled.}
    \label{fig:l2}
\end{figure}

\begin{figure}
    \centering
    \includegraphics[width=0.5\linewidth]{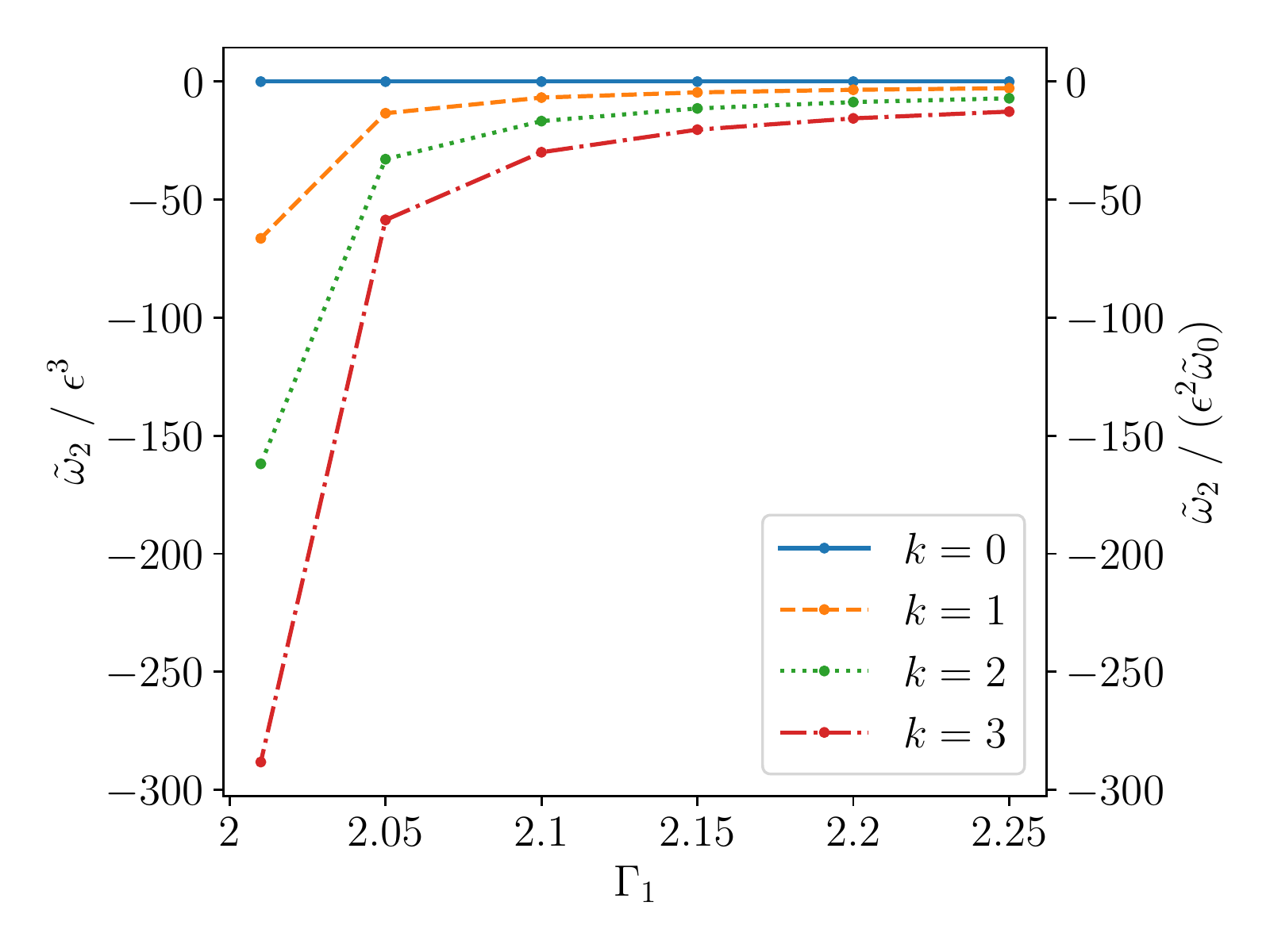}
    \caption{The eigenfrequencies $\tilde{\omega}_2$ of the $(l, m) = (1, 1)$ \textit{r}-modes of the $n = 1$ polytrope with varying $\Gamma_1$. Note that only the $k = 0$ solution remains well behaved in the barotropic limit. The other solutions diverge, implying that $\tilde{\omega}_2$ is promoted to the same order as the leading-order frequency $\tilde{\omega}_0$ and the frequency expansion is spoiled.}
    \label{fig:l1}
\end{figure}

The system of equations~\eqref{eqs:r-mode-non-barotropic} is constrained to stellar models that have $\Gamma_1 \neq \Gamma$. \citep[Indeed,][ was aware of this limitation and so only considered $l = |m|$ \textit{r}-modes for more realistic stellar models in his calculation.]{1982ApJ...256..717S} Although we are unable to consider barotropic stars, we are in a position to explore what happens to the \textit{r}-mode solutions as the stellar model tends towards barotropicity. Motivated by Fig.~\ref{fig:frac}, we will consider the range $2 < \Gamma_1 \leq 2.25$ to approximate $n = 1$ neutron stars. In Figs.~\ref{fig:l3}--\ref{fig:l1}, we show how the eigenfrequencies vary in the $\Gamma_1 \rightarrow \Gamma$ limit. All but the fundamental ($k = 0$) $l = |m|$ eigenfrequencies diverge (and the divergences are worse for the higher overtones). That is, the frequency correction $\tilde{\omega}_2$ grows exponentially as the star becomes more barotropic, showing that care must be taken with the frequency expansion~\eqref{eq:frequencyExpansion} in this limit. If $\tilde{\omega}_2$ becomes comparable in magnitude to $\tilde{\omega}_0$, the frequency will no longer satisfy the Euler equation~\eqref{eq:curl1} at leading order, which in turn will result in a breakdown in the assumed ordering~\eqref{eq:ordering}.

Clearly, for sufficiently small $\epsilon$, $|\tilde{\omega}_2| \ll |\tilde{\omega}_0|$ and the \textit{r}-modes are perfectly well behaved. Indeed, for the more stratified models, the critical rotation at which $\tilde{\omega}_2 \sim O(\tilde{\omega}_0)$ lies well outside the slow-rotation regime~\eqref{eq:slowRotation} where the perturbative framework does not apply. However, as the stellar models become more barotropic, this breakdown sets in at reasonable values of $\epsilon^2 \ll 1$. As illustrative examples, consider the $k = 0$ and $k = 1$ $(l, m) = (2, 2)$ solutions where $\Gamma_1 = 2.01$ (which appear in Fig.~\ref{fig:l2}). The $k = 0$ mode has $\tilde{\omega}_2 / \tilde{\omega}_0 \approx 0.598 \epsilon^2 = O(\epsilon^2)$, whereas $k = 1$ has $\tilde{\omega}_2 / \tilde{\omega}_0 \approx -37.088 \epsilon^2 = O(10 \epsilon^2)$. The $k = 0$ solution will always be valid since, for any slow rotation $\epsilon^2 \ll 1$, $|\tilde{\omega}_2| \ll |\tilde{\omega}_0|$ will be satisfied. This is not so for $k = 1$ where, for (say) $\epsilon^2 \sim 0.1$, the frequency correction approaches $\tilde{\omega}_2 \sim O(\tilde{\omega}_0)$. This feature worsens with higher overtones $k$.

This dependence on $\epsilon$ is important to note since, from the outset, we merely assumed slow rotation according to \eqref{eq:slowRotation}. Now, it is evident that, as we approach barotropicity, the non-fundamental $l = |m|$ \textit{r}-modes only have support at even slower rotation rates. This is an additional constraint on the solutions. In general, if $|\tilde{\omega}_2 / (\epsilon^2 \tilde{\omega}_0)| \gg 1$, then there is an additional dependence on the rotation $\epsilon$ in addition to slow rotation.

Here, we see the competition between the rotation $\epsilon$ of the star and its stratification, parametrised by the adiabatic indices $\Gamma$ and $\Gamma_1$, in supporting \textit{r}-mode oscillations. For rotations of $\epsilon \sim 0.1$ (appropriate for rapidly rotating neutron stars), the frequency corrections of the non-fundamental $l = |m|$ solutions become $\tilde{\omega}_2 \sim O(\epsilon)$ in the $\Gamma_1 \rightarrow \Gamma$ limit. Hence, the leading-order frequencies are no longer simply given by equation~\eqref{eq:omega_0}.

\begin{figure}
\subfloat[$\Gamma_1 = 2.25$]{%
    \includegraphics[width=0.5\columnwidth]{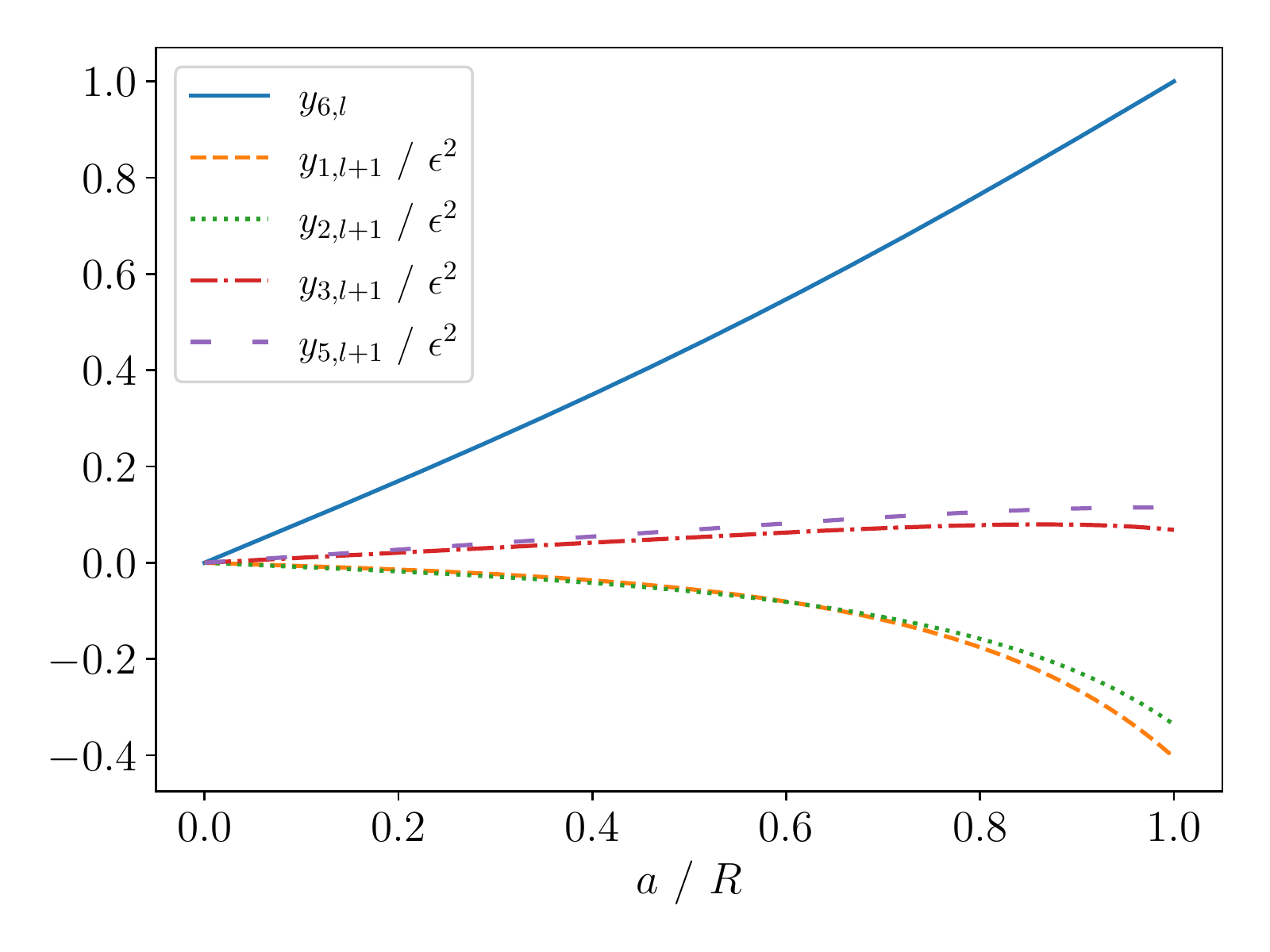}%
}
\subfloat[$\Gamma_1 = 2.01$]{%
    \includegraphics[width=0.5\columnwidth]{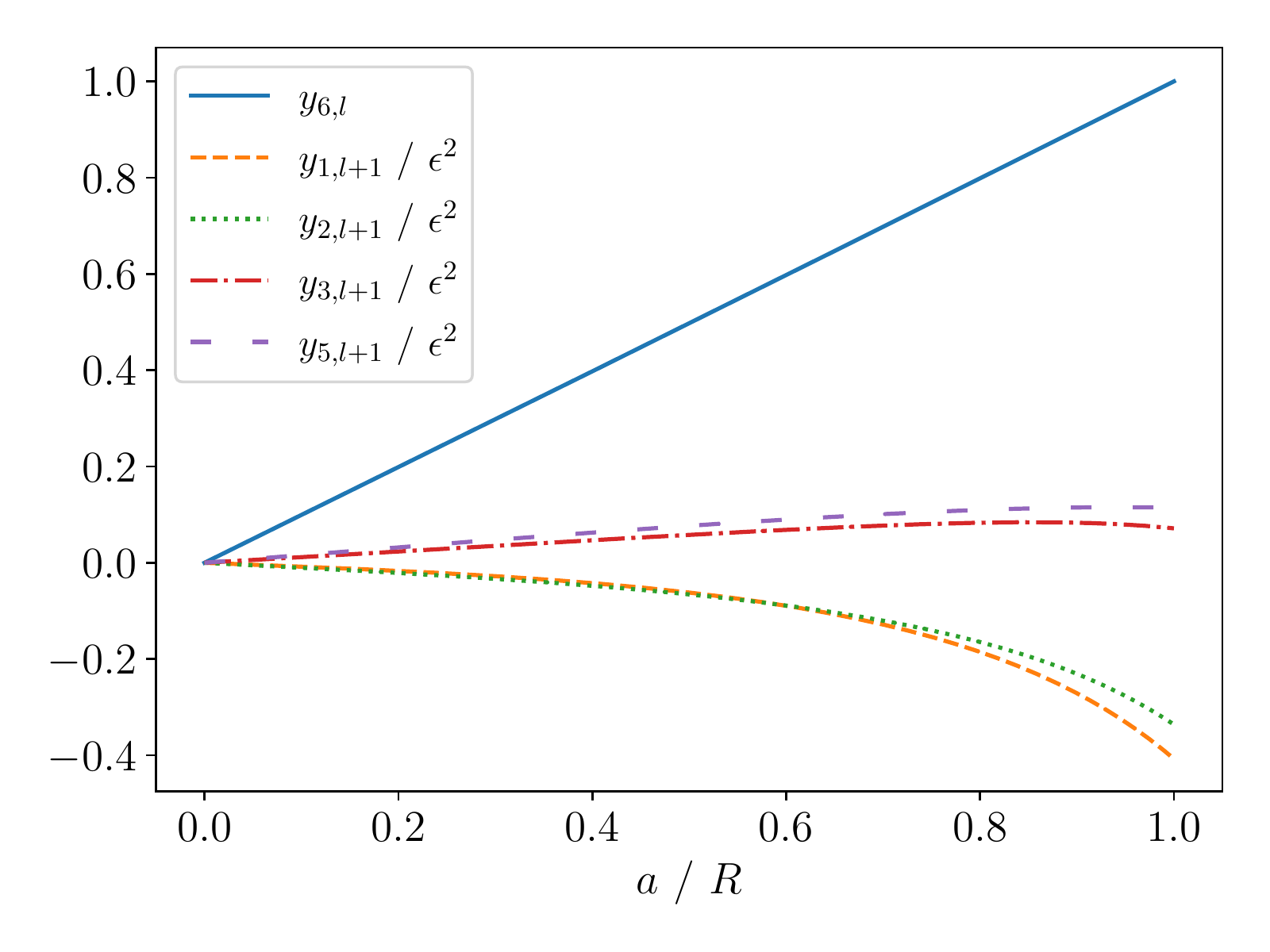}%
}
    \caption{The eigenfunctions of the $(l, m, k) = (2, 2, 0)$ \textit{r}-mode of the $n = 1$ polytrope for different values of $\Gamma_1$. The eigenfunctions of fundamental \textit{r}-modes are relatively unchanged as the star becomes more barotropic.}
    \label{fig:l2m2k0}
\end{figure}

\begin{figure}
\subfloat[$\Gamma_1 = 2.25$]{%
    \includegraphics[width=0.5\columnwidth]{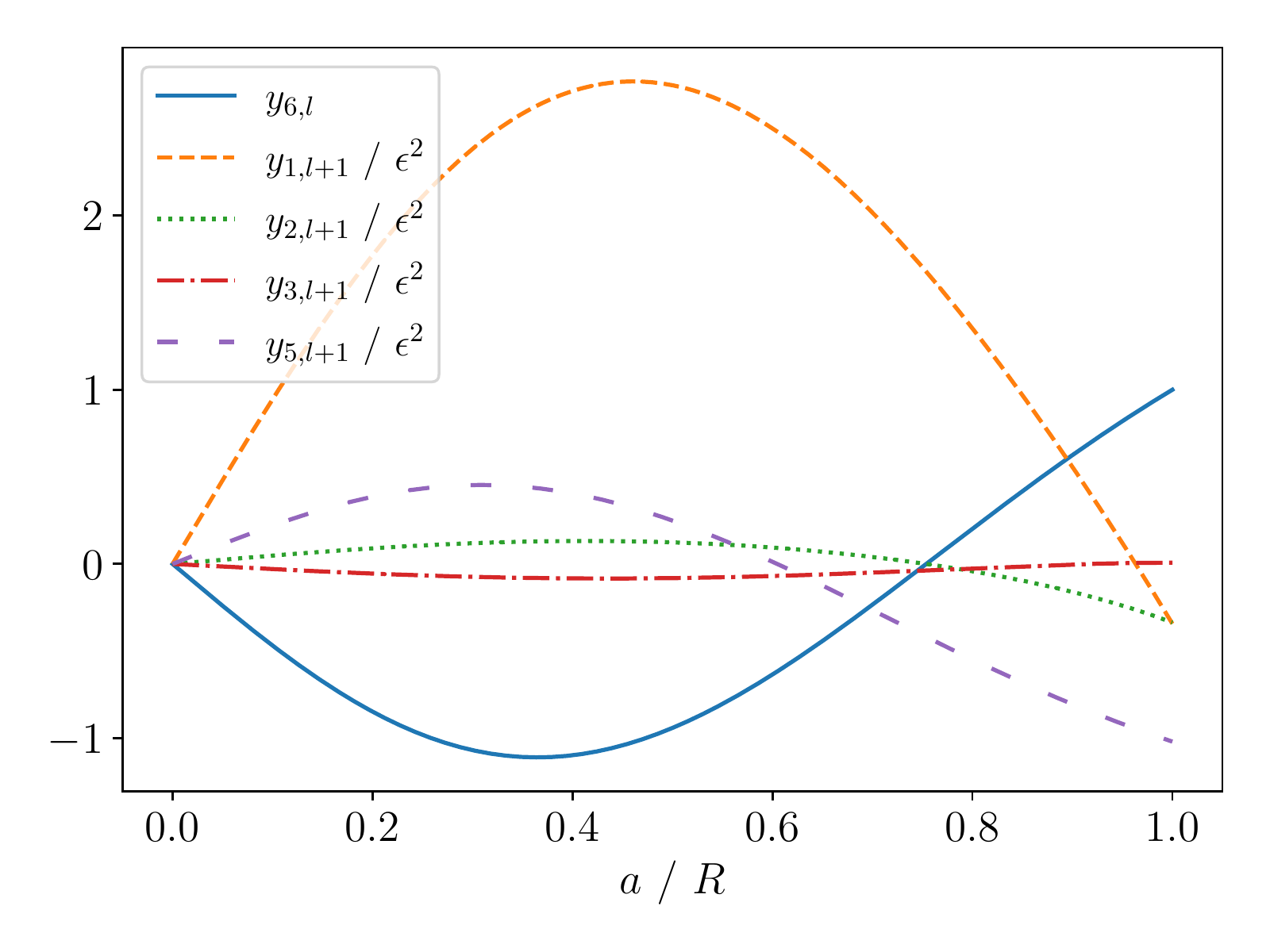}%
}
\subfloat[$\Gamma_1 = 2.01$]{%
    \includegraphics[width=0.5\columnwidth]{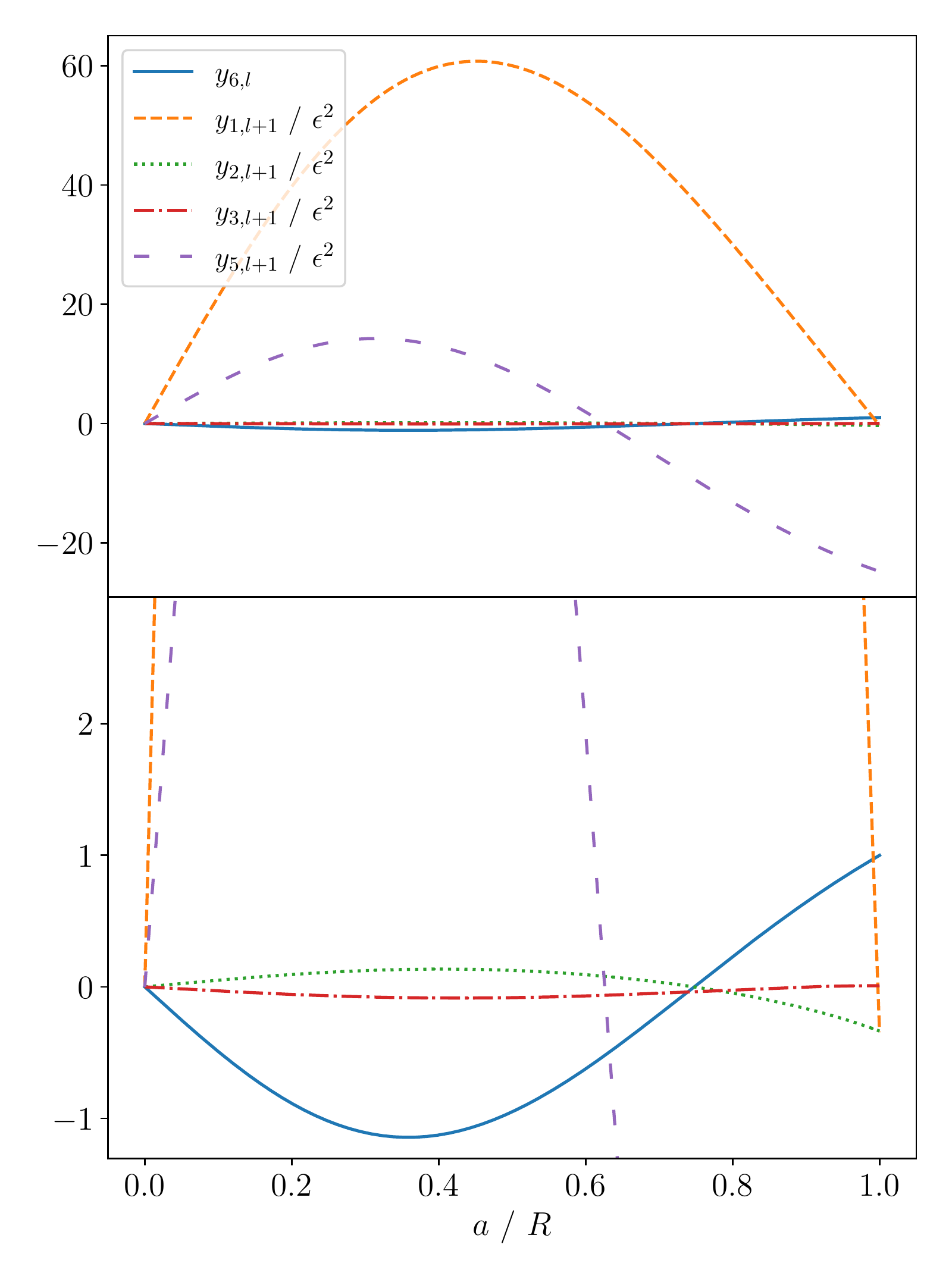}%
}
    \caption{The eigenfunctions of the $(l, m, k) = (2, 2, 1)$ \textit{r}-mode of the $n = 1$ polytrope for different values of $\Gamma_1$. Note the divergences of the polar eigenfunctions $y_{1, l + 1}$ and $y_{5, l + 1}$ in the barotropic limit $\Gamma_1 \rightarrow \Gamma$. This shows how the polar displacement terms are promoted to the same order as the axial displacement $y_{6, l}$ and the perturbation becomes a generic inertial mode on the barotropic star.}
    \label{fig:l2m2k1}
\end{figure}

In Figs.~\ref{fig:l2m2k0} and \ref{fig:l2m2k1}, we present the eigenfunctions for the $k = 0$ and $k = 1$ $(l, m) = (2, 2)$ \textit{r}-modes, respectively. As we witnessed for the eigenfrequencies, the eigenfunctions of the $k = 0$ \textit{r}-mode in Fig.~\ref{fig:l2m2k0} retain their character and are well behaved in the barotropic limit. However, the $k = 1$ overtone has markedly different behaviour, shown in Fig.~\ref{fig:l2m2k1}. As $\Gamma_1 \rightarrow \Gamma$, the polar displacement eigenfunctions $y_{1, l + 1}$ and $y_{5, l + 1}$ diverge, again presenting an issue with the assumed ordering~\eqref{eq:ordering} for moderate rotation rates $\epsilon$. We note that we saw the same qualitative features in all the \textit{r}-modes we calculated, although the divergences had varying levels of severity (as was the case for the eigenfrequencies).

Indeed, for $\epsilon \sim 0.1$, $y_{1, l + 1}$ and $y_{5, l + 1}$ get promoted to $O(1)$ perturbations like $y_{6, l}$ such that the displacement at leading order becomes a mix of polar and axial functions. We show this feature analytically in \citet{other}. The scalar perturbations remain at $O(\epsilon^2)$. This, along with the divergence of the eigenfrequency $\tilde{\omega}_2 \sim O(\epsilon)$, meets the definition of a generic inertial mode.  Therefore, the $k \geq 1$, $l = |m|$ perturbations and the $l > |m|$ perturbations become similar in character to inertial modes as the matter becomes barotropic.

\section{Allowing for local barotropicity}
\label{sec:locally-barotropic}

As we have discussed, equations~\eqref{eqs:r-mode-non-barotropic} assume the star to be globally non-barotropic such that $\Gamma_1 \neq \Gamma$. In reality, neutron stars will have regions where the matter is barotropic (see Fig.~\ref{fig:frac}). We want to explore the extent to which we can drop this assumption and move towards more realistic matter models. We begin with equations~\eqref{eqs:r-mode}, where no assumption about the matter has been made.

One problem immediately rears its ugly head. Without using the algebraic relation~\eqref{eq:relation-y1lminus1}, which has divergent behaviour as $\Gamma_1 \rightarrow \Gamma$, one is unable to decouple the functions $y_{5, l \pm 1}$ that appear in equations~\eqref{eq:continuityPlus}, \eqref{eq:continuityMinus} and \eqref{eq:curl2}. But, given what we found in Section~\ref{sec:non-barotropic}, this may not be too surprising. As a star goes from being globally non-barotropic to globally barotropic, all but the fundamental $l = |m|$ \textit{r}-mode solutions diverge in such a way that they change character and become like generic inertial modes. Should this result hold if the star becomes locally barotropic (even, say, at a point), then it stands to reason that the perturbation equations constructed from the assumed ordering~\eqref{eq:ordering} will not admit any solutions, since no such modes would exist. This seems to be the situation we find ourselves in and we will provide further evidence for this in this section.

As one might expect, the problem simplifies in the $l = |m|$ case. We know that a barotropic star only has one solution for a given $l = |m|$ [the fundamental \textit{r}-mode; see equations~\eqref{eqs:barotrope}]. Therefore, we can examine whether a star that is locally barotropic supports $l = |m|$ \textit{r}-modes with overtones $k \geq 1$.

These modes have no $l - 1$ couplings. Equations~\eqref{eq:divergencePlus} and \eqref{eq:curl2} enable us to remove $y_{5, |m| + 1}$ and $y_{6, |m|}$ from the system of equations. Hence, we obtain the following system for the $l = |m|$ \textit{r}-modes:
\begin{subequations}
\begin{align}
    a \frac{d y_{1, |m| + 1}}{da} &= \left( \frac{V}{\Gamma_1} - 4 - |m| \right) y_{1, |m| + 1} - \left\{ \frac{V}{\Gamma_1} + \frac{|m| + 1}{2 |m| c_1 \epsilon \tilde{\omega}_0} \left[ \frac{(|m| + 1) B}{|m| Q_{|m| + 1}^2} + \frac{3 m}{a^2} \frac{d (a^3 \varepsilon_2)}{da} \right] \right\} y_{2, |m| + 1} + \frac{V}{\Gamma_1} y_{3, |m| + 1}, \\
    a \frac{d y_{2, |m| + 1}}{da} &= a A y_{1, |m| + 1} + (2 + |m| - U - a A) y_{2, |m| + 1} + a A y_{3, |m| + 1}, \\
    a \frac{d y_{3, |m| + 1}}{da} &= (1 - U) y_{3, |m| + 1} + y_{4, |m| + 1}, \\
    a \frac{d y_{4, |m| + 1}}{da} &= - a A U y_{1, |m| + 1} + \frac{U V}{\Gamma_1} y_{2, |m| + 1} + \left[ (|m| + 1) (|m| + 2) - \frac{U V}{\Gamma_1} \right] y_{3, |m| + 1} - U y_{4, |m| + 1}.
\end{align}
\end{subequations}
The behaviour of the eigenfunctions at the centre is given by \eqref{eq:centreFunctions} and the $l + 1$ surface conditions from \eqref{eqs:surface} remain the same.

For our calculation, we use the ratio $\Gamma_1 / \Gamma$ from the BSk19 and BSk21 nuclear-matter models shown in Fig.~\ref{fig:frac}, where $\Gamma$ is obtained from the stellar background. These equations of state depend on the number density of baryons $n_\text{b}$ in the neutron star and therefore introduce dimensionality to the problem. We assume our $n = 1$ stellar model to have $M = 1.4 \ \text{M}_\odot$ and $R = 10 \ \text{km}$. The eigenfrequencies we calculate are listed in Table~\ref{tab:realistic}. We first note that the eigenfrequencies with $k \geq 1$ are larger in magnitude for the BSk19 model than BSk21. This is related to the fact that BSk19 is more weakly stratified and is consistent with our results above with $\Gamma_1 = \text{const}$.

\begin{table}
\centering
\caption{The $l = |m|$ \textit{r}-mode eigenfrequencies $\tilde{\omega}_2 / \epsilon^3$ of an $n = 1$ polytrope, with $M = 1.4 \ \text{M}_\odot$ and $R = 10 \ \text{km}$. The quantity $\Gamma_1 / \Gamma$ is determined from a neutron-star equation of state, BSk19 or BSk21. Note the divergent values with higher overtones.}
\label{tab:realistic}
\begin{tabular}{ c r r r r r r }
\hline
& \multicolumn{2}{c}{$m = 3$} & \multicolumn{2}{c}{$m = 2$} & \multicolumn{2}{c}{$m = 1$} \\ \cline{2-3} \cline{4-5} \cline{6-7}
$k$ & \multicolumn{1}{c}{BSk19} & \multicolumn{1}{c}{BSk21} & \multicolumn{1}{c}{BSk19} & \multicolumn{1}{c}{BSk21} & \multicolumn{1}{c}{BSk19} & \multicolumn{1}{c}{BSk21} \\
\hline
0 & 0.4278 & 0.4282 & 0.3986 & 0.3989 & 0.0000 & 0.0000 \\
1 & $-2.6945$ & $-0.4610$ & $-6.2134$ & $-1.2297$ & $-18.2599$ & $-3.7280$ \\
2 & $-5.1463$ & $-1.6301$ & $-14.6769$ & $-3.9823$ & $-53.6627$ & $-11.9555$ \\
3 & $-11.5551$ & $-3.1305$ & $-28.7989$ & $-7.6752$ & $-94.0447$ & $-23.5897$ \\
\hline
\end{tabular}
\end{table}

\begin{figure}
\subfloat[BSk19]{%
    \includegraphics[width=0.5\columnwidth]{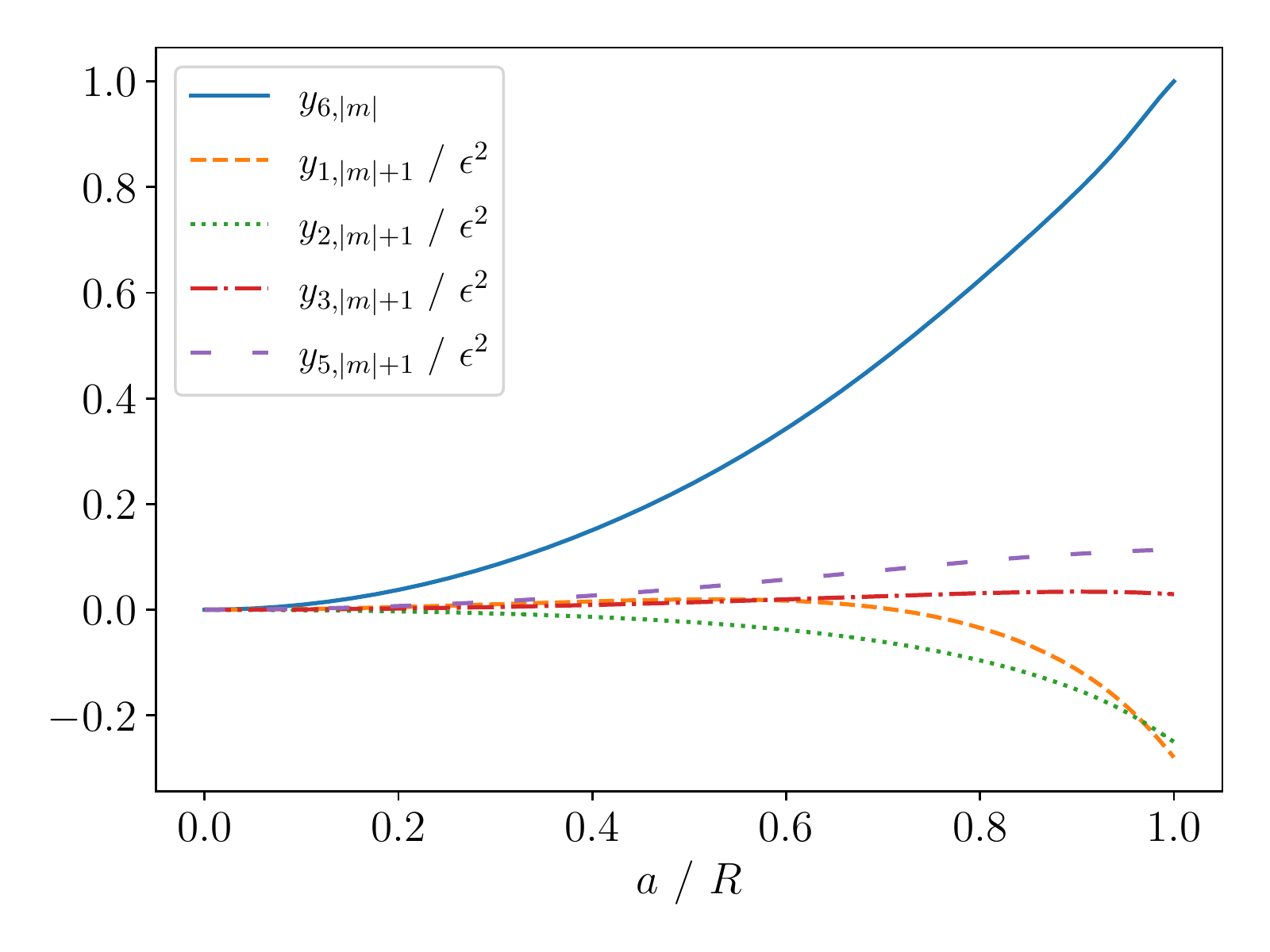}%
}
\subfloat[BSk21]{%
    \includegraphics[width=0.5\columnwidth]{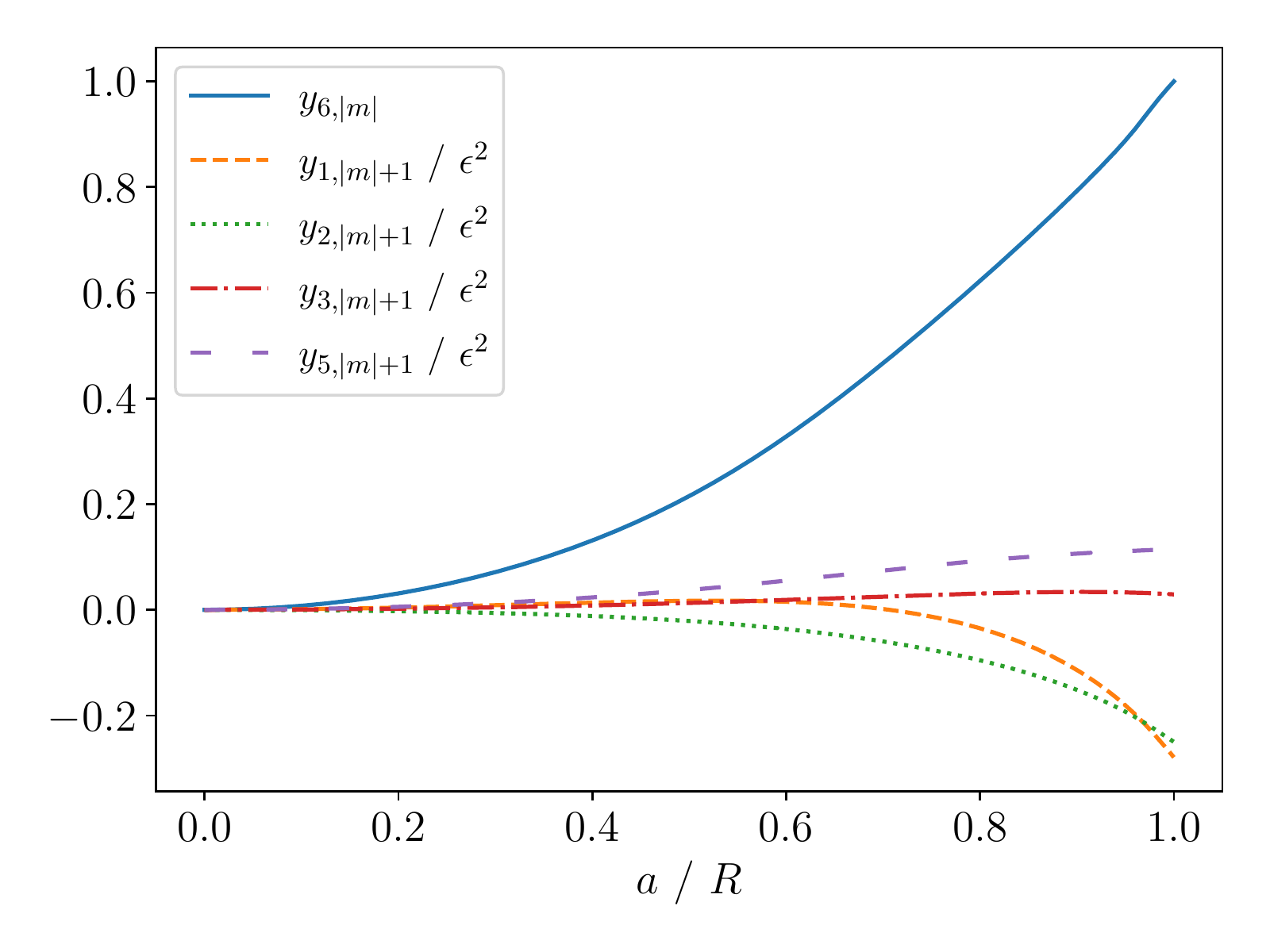}%
}
    \caption{The eigenfunctions of the $(l, m, k) = (|m|, 3, 0)$ \textit{r}-mode of the $n = 1$ polytrope, with $M = 1.4 \ \text{M}_\odot$ and $R = 10 \ \text{km}$, for the BSk19 and BSk21 neutron-star equations of state. Note that the eigenfunctions are similar for the two models.}
    \label{fig:lm3k0}
\end{figure}

\begin{figure}
\subfloat[BSk19]{%
    \includegraphics[width=0.5\columnwidth]{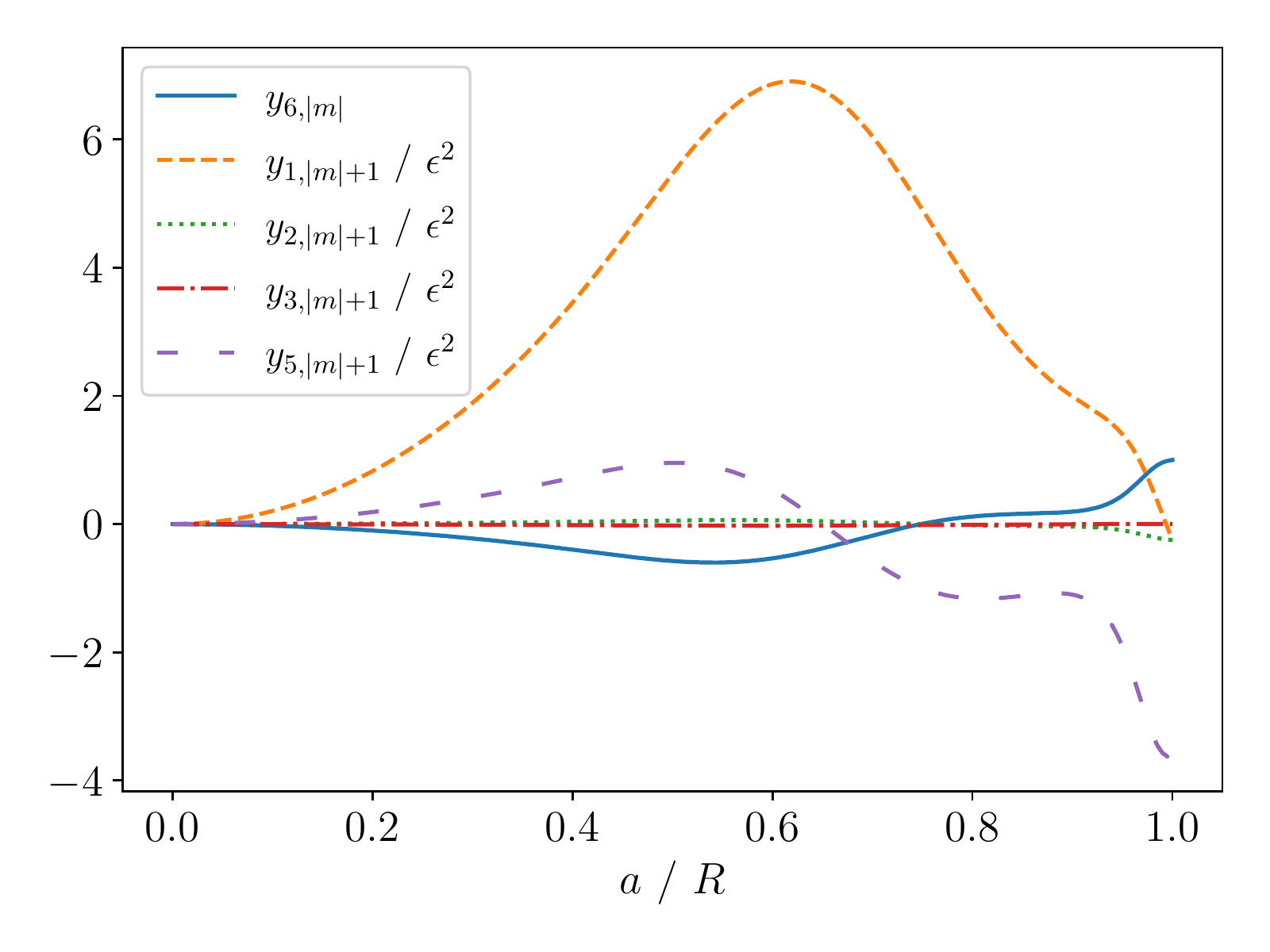}%
}
\subfloat[BSk21]{%
    \includegraphics[width=0.5\columnwidth]{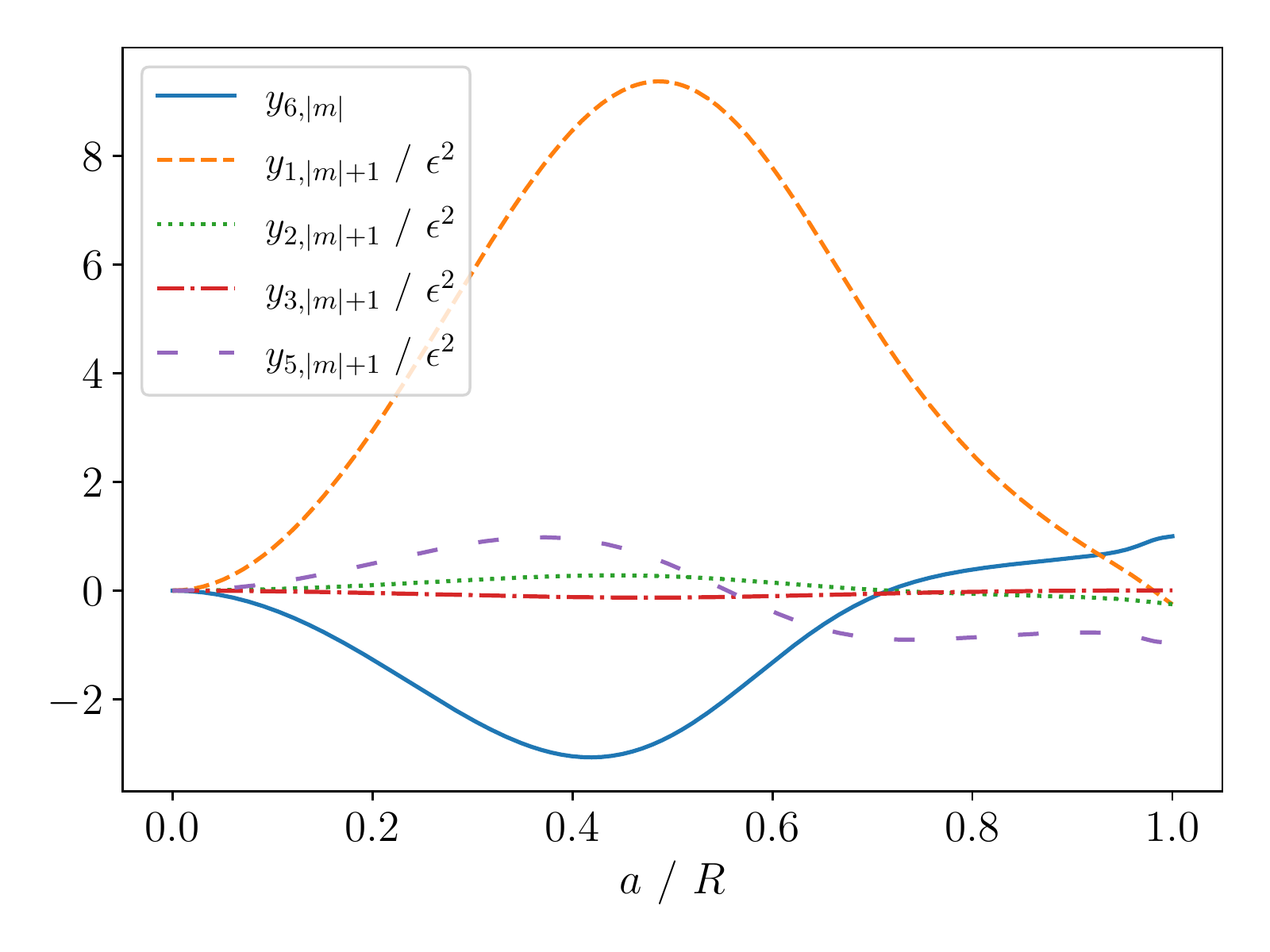}%
}
    \caption{The eigenfunctions of the $(l, m, k) = (|m|, 3, 1)$ \textit{r}-mode of the $n = 1$ polytrope, with $M = 1.4 \ \text{M}_\odot$ and $R = 10 \ \text{km}$, for the BSk19 and BSk21 neutron-star equations of state. Note the large values of the polar displacement eigenfunctions $y_{1, |m| + 1}$ and $y_{5, |m| + 1}$ suggesting that these perturbations are more like generic inertial modes.}
    \label{fig:lm3k1}
\end{figure}

Some of the eigenfrequencies of the BSk19 and BSk21 equations of state do not exhibit particularly strong divergences. As we expect, the $k = 0$ \textit{r}-modes have reasonable values of $\tilde{\omega}_2$. However, we find that all the modes with $k \geq 1$ that we considered have issues with their eigenfunctions. As representative examples, we show the eigenfunctions of the $k = 0$ and $k = 1$ $(l, m) = (|m|, 3)$ \textit{r}-modes in Figs.~\ref{fig:lm3k0} and \ref{fig:lm3k1}, respectively. These two modes do not have strong divergences in $\tilde{\omega}_2$ (see Table~\ref{tab:realistic}). (Although, the $k = 1$ solution will begin to breakdown at spins of $\epsilon^2 \sim 0.1$.) The eigenfunctions of the $k = 0$ \textit{r}-mode in Fig.~\ref{fig:lm3k0} are perfectly well behaved and seem to be relatively insensitive to the linearised equation of state. This is in contrast to the $k = 1$ solution, shown in Fig.~\ref{fig:lm3k1}, which has divergent behaviour in the polar displacement functions $y_{1, |m| + 1}$ and $y_{5, |m| + 1}$. Thus, violating the assumed ordering~\eqref{eq:ordering} at reasonable rates of rotation. To summarise, we do not find any well-behaved solutions with $k \geq 1$; all these solutions have divergences in the $y_{1, |m| + 1}$ and $y_{5, |m| + 1}$ eigenfunctions and many also have divergences in their eigenfrequency $\tilde{\omega}_2$.

Complementary to what we saw for the $\Gamma_1 \neq \Gamma$ calculation in Section~\ref{sec:non-barotropic}, our results for the BSk19 and BSk21 models show that most of the \textit{r}-mode solutions diverge when the star hosts barotropic regions. The only solution that retains its character is the fundamental $l = |m|$ \textit{r}-mode.

\section{Conclusions}
\label{sec:conclusions}

We have considered the role stratification plays in supporting \textit{r}-mode oscillations on slowly rotating, Newtonian stars. We focused on stellar models approximating neutron stars. However, our qualitative results should be the same for other stars that are locally barotropic.

In using the linearised equations derived by \citet{1982ApJ...256..717S}, which are only valid for stellar models that are globally non-barotropic $\Gamma_1 \neq \Gamma$, we found that the majority of solutions exhibited divergences in the barotropic limit $\Gamma_1 \rightarrow \Gamma$. These divergences occur for the $k \geq 1$, $l = |m|$ \textit{r}-modes and the $l > |m|$ \textit{r}-modes and manifest themselves in two ways. (1) The frequency correction $\tilde{\omega}_2$ becomes comparable in magnitude to the linear term $\tilde{\omega}_0$ such that $\tilde{\omega}_2 / \tilde{\omega}_0 \sim O(1)$ for modest rates of rotation $\epsilon$. This spoils the usual identification of the \textit{r}-mode frequency at leading order. (2) The polar terms in the displacement vector increase to $O(1)$; the same order as the axial term. Thus, the displacement becomes a mixture of polar and axial functions at zeroth order in rotation. These two divergences lead to the solutions becoming generic inertial modes. However, as one may expect, none of these features are seen in the fundamental ($k = 0$) $l = |m|$ \textit{r}-mode solutions, which remain well behaved in the non-stratified limit.

Moving beyond the globally non-barotropic approximation, we considered stars that have varying stratification with barotropic regions. The perturbation equations cannot be expressed as a standard eigenvalue problem for general $l \geq |m|$ \textit{r}-modes if the star is barotropic at some point. However, the situation becomes tractable for $l = |m|$ modes. We calculated the $l = |m|$ \textit{r}-modes of an $n = 1$ polytrope with the perturbations described by the BSk19 and BSk21 nuclear-matter equations of state. Numerically, we obtained solutions in addition to the fundamental \textit{r}-mode with $k \geq 1$. Although these numerical solutions exist, they also present divergent behaviour in $\tilde{\omega}_2$ and the polar displacement functions. Formally, these results confirm the expectation that for stratified stars there will be a critical rotation rate above which the fluid only supports the fundamental $l = |m|$ \textit{r}-modes. In addition, we have shown that the same is true for stars that are barotropic in a local region. The remaining Coriolis-driven perturbations must join the general inertial-mode family. Furthermore, depending on how stratified the equation of state is, the rotation at which the solutions change character can be very modest indeed. This implies that rapidly rotating neutron stars, which are of interest for gravitational-wave observations, may only host the fundamental $l = |m|$ \textit{r}-modes. However, in order to verify which modes persist at fast rotations, we must address the issues with the relativistic problem. Only then can we construct realistic neutron-star models using nuclear-matter equations of state.

This work has some natural extensions. Our analysis was limited to examine perturbations that have the \textit{r}-mode ordering in a slow-rotation expansion. Clearly, the divergences we have witnessed spoil this ordering and the equations we use are not strictly valid when they arise. It would therefore be interesting to examine the behaviour of the divergent \textit{r}-modes using machinery appropriate for the inertial modes. It is notable that the vast majority of the work on  inertial modes has focused on barotropic stars. Moreover, efforts in the non-barotropic context tend to assume $\Gamma_1 = \text{const}$ \citep{1987MNRAS.224..513L,2000ApJS..129..353Y}, with \citet{2005PhRvD..71h3001V} and \citet{2022PhRvD.106j3009K} as exceptions. Further work should consider whether realistic stratification has important consequences for the \textit{r}-modes of relativistic stars.

\section*{Acknowledgements}

The authors are grateful for support from STFC via grant number ST/V000551/1. The contribution of FG was partly carried out at the Institute for Nuclear Theory at the University of Washington during the ``Neutron Rich Matter on Heaven and Earth'' workshop, which is supported by the U.S. Department of Energy grant DE-FG02-00ER41132. The contribution of NA was partly carried out at the Aspen Center for Physics, which is supported by National Science Foundation grant PHY-1607611. He also thanks the Simons Foundation for generous travel support.

\section*{Data availability}

Additional data underlying this article will be shared on reasonable request.


\bibliographystyle{mnras}
\bibliography{numerical-bibliography}


\appendix

\section{Second-order axial perturbations}
\label{app:axial}

We expand the axial perturbation as 
\begin{equation}
    U_{l'} = U_{l', 0} + U_{l', 2} + O(\epsilon^4),
\end{equation}
where $U_{l', 0} = O(1)$ and $U_{l', 2} = O(\epsilon^2)$. We develop the angular components of the Euler equation~\eqref{eq:perturbedEuler} further to $O(\epsilon^4)$ and find 
\begin{equation}
\begin{split}
    0 = \sum_l \bigg( &2 \Omega \omega_0 [2 W_l - l (l + 1) V_l] \cos \theta \, Y_l^m + 2 \Omega \omega_0 (W_l - V_l) \sin \theta \, \partial_\theta Y_l^m \\
    &+ \big\{ 2 [l (l + 1) \omega_0 - m \Omega] \omega_2 - [l (l + 1) \omega_0 - 8 m \Omega] \omega_0 \varepsilon_2 \big\} U_{l, 0} Y_l^m \\
    &+ [l (l + 1) \omega_0 - 2 m \Omega] \omega_0 U_{l, 2} Y_l^m \\
    &+ 3 [l (l + 1) \omega_0 - 12 m \Omega] \omega_0 \varepsilon_2 U_{l, 0} \cos^2 \theta \, Y_l^m + 6 \omega_0^2 \varepsilon_2 U_{l, 0} \cos \theta \sin \theta \, \partial_\theta Y_l^m \bigg).
\end{split}
\end{equation}
By the recursion relations~\eqref{eqs:recursion} with the orthogonality of the spherical harmonics~\eqref{eq:orthogonality}, we have
\begin{equation}
\begin{split}
    0 = 2& (l' + 1) Q_{l'} \Omega \omega_0 [W_{l' - 1} - (l' - 1) V_{l' - 1}] - 2 l' Q_{l' + 1} \Omega \omega_0 [W_{l' + 1} + (l' + 2) V_{l' + 1}] \\
    &+ \big\{ 2 [l' (l' + 1) \omega_0 - m \Omega] \omega_2 - [l' (l' + 1) \omega_0 - 8 m \Omega] \omega_0 \varepsilon_2 + 3 (Q_{l' + 1}^2 + Q_{l'}^2) [l' (l' + 1) \omega_0 - 12 m \Omega] \omega_0 \varepsilon_2 \\
    &\qquad+ 6 [l' Q_{l' + 1}^2 - (l' + 1) Q_{l'}^2] \omega_0^2 \varepsilon_2 \big\} U_{l', 0} \\
    &+ [l' (l' + 1) \omega_0 - 2 m \Omega] \omega_0 U_{l', 2} \\
    &+ 3 Q_{l' - 1} Q_{l'} [(l' - 2) (l' + 1) \omega_0 - 12 m \Omega] \omega_0 \varepsilon_2 U_{l' - 2, 0} + 3 Q_{l' + 2} Q_{l' + 1} [l' (l' + 3) \omega_0 - 12 m \Omega] \omega_0 \varepsilon_2 U_{l' + 2, 0}.
\end{split}
\end{equation}
Clearly, for $l' = l$, the term with $U_{l', 2}$ will vanish due to equation~\eqref{eq:omega_0}. Although we note that in principle $U_{l, 2} \neq 0$. Then we need only work with the $O(1)$ contribution to $U_l$. We have 
\begin{equation}
\begin{split}
    0 = 2& (l + 1) Q_l \Omega \omega_0 [W_{l - 1} - (l - 1) V_{l - 1}] - 2 l Q_{l + 1} \Omega \omega_0 [W_{l + 1} + (l + 2) V_{l + 1}] \\
    &+ \big\{ 2 [l (l + 1) \omega_0 - m \Omega] \omega_2 - [l (l + 1) \omega_0 - 8 m \Omega] \omega_0 \varepsilon_2 + 3 (Q_{l + 1}^2 + Q_l^2) [l (l + 1) \omega_0 - 12 m \Omega] \omega_0 \varepsilon_2 \\
    &\qquad+ 6 [l Q_{l + 1}^2 - (l + 1) Q_l^2] \omega_0^2 \varepsilon_2 \big\} U_{l, 0}.
\end{split}
\end{equation}
This is the dimensional form of equation~\eqref{eq:curl2}. We see that we may obtain the $l \pm 2$ second-order axial corrections by examining $l' = l + 2$,
\begin{subequations}
\begin{equation}
    0 = 2 (l + 3) Q_{l + 2} \Omega \omega_0 [W_{l + 1} - (l + 1) V_{l + 1}] + [(l + 2) (l + 3) \omega_0 - 2 m \Omega] \omega_0 U_{l + 2, 2} + 3 Q_{l + 1} Q_{l + 2} [l (l + 3) \omega_0 - 12 m \Omega] \omega_0 \varepsilon_2 U_{l, 0},
\end{equation}
and $l' = l - 2$,
\begin{equation}
    0 = - 2 (l - 2) Q_{l - 1} \Omega \omega_0 (W_{l - 1} + l V_{l - 1}) + [(l - 2) (l - 1) \omega_0 - 2 m \Omega] \omega_0 U_{l - 2, 2} + 3 Q_l Q_{l - 1} [(l - 2) (l + 1) \omega_0 - 12 m \Omega] \omega_0 \varepsilon_2 U_{l, 0}.
\end{equation}
\end{subequations}
These couplings were also discussed by \citet{1981Ap&SS..78..483S}.


\bsp
\label{lastpage}
\end{document}